\documentclass[a4paper,11pt]{article}
\pdfoutput=1 
\usepackage{aas_macros,braket}
\usepackage{jcappub,natbib} 
\usepackage[T1]{fontenc} 

\title{Statistical anisotropy in galaxy ellipticity correlations}

\author[a]{Maresuke Shiraishi,}
\author[b,c]{Teppei Okumura,}
\author[d]{and Kazuyuki Akitsu}

\affiliation[a]{School of General and Management Studies, Suwa University of Science, Chino, Nagano 391-0292, Japan}
\affiliation[b]{Academia Sinica Institute of Astronomy and Astrophysics (ASIAA), No. 1, Section 4, Roosevelt Road, Taipei 10617, Taiwan}
\affiliation[c]{Kavli Institute for the Physics and Mathematics of the Universe (WPI), UTIAS, The University of Tokyo, Chiba 277-8583, Japan}
\affiliation[d]{School of Natural Sciences, Institute for Advanced Study, 1 Einstein Drive, Princeton, NJ 08540, USA}

\emailAdd{shiraishi\_maresuke@rs.sus.ac.jp}
\emailAdd{tokumura@asiaa.sinica.edu.tw}
\emailAdd{kakitsu@ias.edu}

\abstract{
  As well as the galaxy number density and peculiar velocity, the galaxy intrinsic alignment can be used to test the cosmic isotropy. We study distinctive impacts of the isotropy breaking on the configuration-space two-point correlation functions (2PCFs) composed of the spin-2 galaxy ellipticity field. For this purpose, we build a formalism for general types of the isotropy-violating 2PCFs and a methodology to efficiently compute them by generalizing the polypolar spherical harmonic decomposition approach to the spin-weighted version. As a demonstration, we analyze the 2PCFs when the matter power spectrum has a well-known $g_*$-type isotropy-breaking term (induced by, e.g., dark vector fields). We then confirm that some anisotropic distortions indeed appear in the 2PCFs and their shapes rely on a preferred direction causing the isotropy violation, $\hat{d}$. Such a feature can be a distinctive indicator for testing the cosmic isotropy. Comparing the isotropy-violating 2PCFs computed with and without the plane parallel (PP) approximation, we find that, depending on $\hat{d}$, the PP approximation is no longer valid when an opening angle between the directions towards target galaxies is ${\cal O}(1^\circ)$ for the density-ellipticity and velocity-ellipticity cross correlations and around $10^\circ$ for the ellipticity auto correlation. This suggests that an accurate test for the cosmic isotropy requires the formulation of the 2PCF without relying on the PP approximation.
}

\begin{document}



\maketitle
\flushbottom

\section{Introduction}

Global isotropy of the Universe is a major conjecture in cosmology, and it has been supported by various types of the cosmic observations so far. However, the possibility of a small isotropy violation still has been allowed and forthcoming observations will reveal it.

Theoretically, the statistical isotropy of the Universe can be violated by the existence of strong anisotropic sources like vector fields. There are already various inflationary scenarios including vector fields motivated by, e.g., magnetogenesis and axiverse models (see e.g. refs.~\cite{Dimastrogiovanni:2010sm,Soda:2012zm,Maleknejad:2012fw} for review). Behaviors of vector fields as dark matter and dark energy candidates have also been thoroughly argued (e.g. refs.~\cite{BeltranJimenez:2008iye,Hambye:2008bq,Graham:2015rva,Bastero-Gil:2018uel}). Testing the statistical isotropy in cosmic observables therefore becomes a powerful diagnostic approach of such scenarios. The cosmic microwave background (CMB) observations have already tightly constrained the statistical isotropy breaking \cite{Planck:2018jri,Planck:2019evm,Planck:2019kim}, and a bit weaker limits have been obtained via the measurement of the galaxy number density from galaxy surveys \cite{Pullen:2010zy,Sugiyama:2017ggb}.

Regarding observables in the galaxy surveys, not only conventional spin-0 fields that are number density and peculiar velocity,%
\footnote{ 
The peculiar velocity can also have a vorticity-induced spin-1 component, however, it is negligibly small in the standard cosmology.
}
but also a spin-2 field, ellipticity, have rich information and hence the galaxy ellipticity field has recently come into use as a beneficial cosmological probe (e.g. refs.~\citep{Chisari:2013dda,Schmidt:2015xka,Chisari:2016xki,Kogai:2018nse,Okumura:2019ozd,Taruya:2020tdi,Kogai:2020vzz,Akitsu:2020jvx,Okumura:2021xgc,Saga:2022frj,Okumura:2023pxv,Kurita:2023qku}). Implementation of the ellipticity field in the isotropy test is naturally expected to improve the constraint or yield some novel information. Motivated by this, in this paper, we, for the first time, study distinctive impacts of the isotropy breaking on the configuration-space two-point correlation functions (2PCFs) composed of the ellipticity field. For this purpose, we develop a formalism for general types of the isotropy-violating 2PCFs and a methodology to efficiently compute them.

The technique of spin-weighted polypolar spherical harmonic (PolypoSH) decomposition can be used as a powerful tool to compute galaxy statistics including the intricate spin and angular dependence.%
\footnote{
For different but similar approaches, see refs.~\cite{Vlah:2019byq,Vlah:2020ovg,Matsubara:2022ohx,Matsubara:2022eui}. 
}
Using this technique, previous studies presented the analysis of wide-angle effects of spin-0 \citep{Szalay:1997cc,Szapudi:2004gh,Yoo:2013zga,Castorina:2017inr,Taruya:2019xsf,Castorina:2019hyr,Shiraishi:2020nnw,Shiraishi:2021oau} and higher-spin \cite{Shiraishi:2020vvj} field correlations, and isotropy-breaking signatures of spin-0 field ones \cite{Shiraishi:2016wec,Bartolo:2017sbu,Akitsu:2019avy,Shiraishi:2020pea}. We here generalize the decomposition technique to deal with general types of isotropy-breaking signatures on higher-spin field correlations. 

As a numerical application of this new methodology, we analyze the 2PCFs generated in the case where the matter power spectrum has a well-known $g_*$-type isotropy-breaking term \cite{Ackerman:2007nb,Pullen:2010zy} induced by, e.g., dark vector fields (hereinafter called the $g_*$ model). For the first step, we examine the plane-parallel (PP) limit, where an opening angle (dubbed as $\Theta$) between two line-of-sight (LOS) directions toward the positions of galaxies, $\hat{x}_1 \equiv {\bf x}_1 / x_1$ and $\hat{x}_2 \equiv {\bf x}_2 / x_2$, is small enough that we can approximate $\hat{x}_1 \simeq \hat{x}_2 = \hat{x}_{\rm p}$. Hence, the 2PCFs are computable using the spin-weighted bipolar spherical harmonic (BipoSH) basis, $\{Y_{\ell}(\hat{x}_{12}) \otimes {}_{\lambda'}Y_{\ell'}(\hat{x}_{\rm p})\}_{LM}$, where $\hat{x}_{12} \equiv ({\bf x}_1 - {\bf x}_2) / |{\bf x}_1 - {\bf x}_2|$ is the direction of the separation vector between the positions of target galaxies. Showing obtained 2PCF signals as a function of parallel and perpendicular elements of ${\bf x}_{12}$ as in ref.~\cite{Okumura:2019ned}, we find some distinctive distortions due to the isotropy breaking, which could be a key indicator for testing the cosmic isotropy.

The PP approximation would not be applicable to the analysis in futuristic wider galaxy surveys. Therefore, as the next step, we treat $\hat{x}_1$ and $\hat{x}_2$ separately, and compute the 2PCFs by introducing the spin-weighted tripolar spherical harmonic (TripoSH) basis $\{ Y_{\ell}(\hat{x}_{12}) \otimes \{ {}_{\lambda_1} Y_{\ell_1}(\hat{x}_1) \otimes {}_{\lambda_2}Y_{\ell_2}(\hat{x}_2) \}_{\ell'} \}_{LM}$. Comparing the exact results with the PP-limit ones for the $g_*$ model, we measure the error level of the PP approximation as a function of $\Theta$. We then find that it is sensitive to a global preferred direction causing the isotropy violation and exceeds $10\%$ up to $\Theta = {\cal O}(1^\circ)$ at worst, indicating the importance of beyond the PP-limit analysis for testing the cosmic isotropy more accurately.

This paper is organized as follows. In the next section, we derive a formalism on the galaxy density, velocity and ellipticity fields originating from a general type of the isotropy-breaking matter fluctuation. In section~\ref{sec:PolypoSH}, we build an efficient computation methodology for general types of the isotropy-breaking 2PCFs with and without the PP approximation by means of the BipoSH and TripoSH techniques, respectively. Section~\ref{sec:result} presents a numerical analysis of the 2PCFs in the $g_*$ model, and the final section concludes this work.

\section{Galaxy density, velocity and ellipticity fields induced by a general type of the isotropy-breaking matter fluctuation} \label{sec:linear}

For later analysis of the large-scale statistics, in this section, we derive a linear theory expressions of the galaxy density, velocity and ellipticity fields when the underlying matter fluctuation breaks isotropy in a general way.

\subsection{Isotropy-breaking matter fluctuation}

In the following analysis, we impose the statistical homogeneity of the real-space matter fluctuation $\delta_{\rm m}$; thus, the matter power spectrum generally takes the form
\begin{align}
  \Braket{\delta_{\rm m}({\bf k}_1) \delta_{\rm m}({\bf k}_2)} = (2\pi)^3 \delta^{(3)}({\bf k}_1 + {\bf k}_2) P_{\rm m}({\bf k}_1) . \label{eq:Pm}
\end{align}
As we are in position to consider the statistical isotropy violation of $\delta_{\rm m}$, the $\hat{k}$ dependence remains in $P_{\rm m}({\bf k})$. We also assume the Gaussianity of $\delta_{\rm m}$ and therefore may ignore any impact of higher-order statistics. Here, although $\delta_{\rm m}$ depends on time, redshift or the comoving distance, it is not explicitly stated as an argument for notational convenience. This convention is adapted to all variables henceforth unless the parameter dependence is nontrivial and the explicit representation is needed.

Without loss of generality, we can expand $P_{\rm m}({\bf k})$ according to
\begin{align}
  P_{\rm m}({\bf k}) &= \bar{P}_{\rm m}(k)
  \left[ 1 + \sum_{L>0} \sum_M G_{L M}(k) Y_{LM}(\hat{k}) \right] \nonumber \\
 &= \bar{P}_{\rm m}(k) \sum_{LM} G_{L M}(k) Y_{LM}(\hat{k}) , \label{eq:Pm_GLM}
\end{align}
where $G_{LM}^* = (-1)^{M} G_{L, -M}$ and $G_{L={\rm odd}, M} = 0$. Breaking the statistical isotropy gives rise to nonvanishing $G_{L > 0, M}$. On the other hand, we define $G_{00} \equiv \sqrt{4\pi}$, so that $P_{\rm m}({\bf k}) = \bar{P}_{\rm m}(k)$ holds if $\delta_{\rm m}$ is statistically isotropic. Because of observational bounds as $|G_{L> 0, M}| \ll 1$, the matter density field satisfying eq.~\eqref{eq:Pm_GLM} takes the form
\begin{align}
  \delta_{\rm m}({\bf k}) = \bar{\delta}_{\rm m}({\bf k})
  \left[ 1 + \frac{1}{2} \sum_{L > 0} \sum_{M} G_{L M}(k) Y_{LM}(\hat{k}) \right]
  , \label{eq:delm_GLM}
\end{align}
where $\bar{\delta}_{\rm m}$ denotes the isotropy-conserving part in the matter fluctuation, whose power spectrum is given by
\begin{align}
  \Braket{\bar{\delta}_{\rm m}({\bf k}_1) \bar{\delta}_{\rm m}({\bf k}_2)} = (2\pi)^3 \delta^{(3)}({\bf k}_1 + {\bf k}_2) \bar{P}_{\rm m}(k_1) . \label{eq:Pm_iso}
\end{align}

Let us also introduce another mathematically-equivalent representation:
\begin{align}
    P_{\rm m}({\bf k}) &= \bar{P}_{\rm m}(k)
    \left[ 1
      + {\cal G}_{ij}^{(2)}(k) \hat{k}_{i} \hat{k}_{j}
      + {\cal G}_{ijkl}^{(4)}(k) \hat{k}_{i} \hat{k}_{j} \hat{k}_{k} \hat{k}_{l}
      + {\cal G}_{ijklmn}^{(6)}(k) \hat{k}_{i} \hat{k}_{j} \hat{k}_{k} \hat{k}_{l} \hat{k}_{m} \hat{k}_{n}  
      + \cdots  \right] \nonumber \\
  &= \bar{P}_{\rm m}(k)
  \left[1 +  \sum_{L > 0} {\cal G}_{i_1 \cdots i_L}^{(L)}(k) \prod_{r=1}^L \hat{k}_{i_r} \right] , \label{eq:Pm_calG}
\end{align}
where ${\cal G}_{i_1 \cdots i_L}^{(L)}$ is a totally symmetric rank-$L$ traceless tensor field obeying ${\cal G}_{i_1 \cdots i_L}^{(L)} \in \mathbb{R}$ and ${\cal G}_{i_1 \cdots i_L}^{(L = \rm odd)} = 0$. In this expression, nonvanishing ${\cal G}_{i_1 \cdots i_L}^{(L)}$ fully characterize isotropy-breaking signatures. Observational bounds as $|{\cal G}_{i_1 \cdots i_L}^{(L)}| \ll 1$ lead to 
\begin{align}
  \delta_{\rm m}({\bf k}) = \bar{\delta}_{\rm m}({\bf k})
  \left[ 1 + \frac{1}{2} \sum_{L > 0} {\cal G}_{i_1 \cdots i_L}^{(L)}(k) \prod_{r=1}^L \hat{k}_{i_r} \right] . \label{eq:delm_calG}
\end{align}

In the remainder of this section, we first derive the linear theory formulae with the later tensorial representation, and finally rewrite them using the former harmonic one for good compatibility with the PolypoSH decomposition performed in the next section. We then utilize the conversion formulae: 
\begin{align}
  \begin{split}
  G_{LM}(k) &= \int d^2 \hat{k} \, Y_{LM}^*(\hat{k}) {\cal G}_{i_1 \cdots i_L}^{(L)}(k) \prod_{r=1}^L \hat{k}_{i_r} , \\
  {\cal G}_{i_1 \cdots i_L}^{(L)}(k)
  &= \frac{(2L+1)!!}{4\pi L!} \int d^2 \hat{k} \left( \prod_{r=1}^L \hat{k}_{i_r} \right) \sum_{M} G_{L M}(k) Y_{LM}(\hat{k}), \label{eq:GLM_2_calG}
  \end{split}
\end{align}
where the latter one has been derived employing eq.~\eqref{eq:math_intvec}.

\subsection{Spin-0 galaxy field: density $\delta$ and velocity $u$}

Here we formulate the number density fluctuation, $\delta({\bf x}) \equiv n({\bf x}) / n_{\rm av}(x) - 1$, and the LOS component of peculiar velocity, $u({\bf x}) \equiv {\bf v}({\bf x}) \cdot \hat{x}$, in the redshift space.

As shown in eq.~\eqref{eq:delm_calG}, the real-space matter density field $\delta_{\rm m}$ is anisotropically modulated by some extra tensor field ${\cal G}_{i_1 \cdots i_L}^{(L)}$ (and hence the isotropy-breaking matter power spectrum \eqref{eq:Pm_calG} is realized). Such a field could also give characteristic impacts on the bias relation between galaxy and matter number densities \cite{Schmidt:2015xka,Kogai:2020vzz}. In a similar manner to refs.~\cite{Schmidt:2015xka,Assassi:2015fma,Kogai:2020vzz}, we fully expand the real-space galaxy number density field $\delta^r$ with respect to $\frac{\partial_i \partial_j}{\partial^2} \delta_{\rm m}$ by using the Kronecker delta $\delta_{ij}^{\rm K}$ and ${\cal G}_{i_1 \cdots i_L}^{(L)}$ for tensor contractions. We then find that $\delta^r$ at leading order of $\delta_{\rm m}$ is expressed with two different bias parameters $b_{\rm g}$ and $b_{\rm g}^{(2)}$; namely,
\begin{align}
  \delta^r({\bf x})
  = \int \frac{d^3 k}{(2\pi)^3} e^{i {\bf k} \cdot {\bf x}}
  \left[b_{\rm g} +  \frac{1}{2}b_{\rm g}^{(2)} {\cal G}_{ij}^{(2)}(k) \hat{k}_i \hat{k}_j \right]
  \delta_{\rm m}({\bf k}) .
\end{align}
Here, $b_{\rm g}$ is equivalent to a linear bias parameter introduced in standard isotropy-conserving universe models, while $b_{\rm g}^{(2)}$ is a newly-introduced one due to nonvanishing ${\cal G}_{ij}^{(2)}$, or equivalently, nonvanishing $G_{2M}$. If moving to the redshift space, the usual distortion terms are added in this expression. Evaluation up to linear order of ${\cal G}_{i_1 \cdots i_L}^{(L)}$ with eq.~\eqref{eq:delm_calG} yields
\begin{align}
\begin{split}
  \delta({\bf x})
  &= \bar{\delta}({\bf x}) +  \delta_{\rm ani}^{\rm std}({\bf x}) + \delta_{\rm ani}^{\rm new}({\bf x}) , \\
  \bar{\delta}({\bf x}) &\equiv \int \frac{d^3 k}{(2\pi)^3} e^{i {\bf k} \cdot {\bf x}} 
  \left[ b_{\rm g}
  - i \frac{\alpha f}{kx} (\hat{k} \cdot \hat{x})  
  + f \, (\hat{k} \cdot \hat{x})^2  \right]
  \bar{\delta}_{\rm m}({\bf k}) , \\
  \delta_{\rm ani}^{\rm std}({\bf x}) &\equiv \int \frac{d^3 k}{(2\pi)^3} e^{i {\bf k} \cdot {\bf x}}
  \left[ b_{\rm g} - i \frac{\alpha f}{kx} (\hat{k} \cdot \hat{x})  
    + f \, (\hat{k} \cdot \hat{x})^2 \right]
  \left[ \frac{1}{2} \sum_{L > 0} {\cal G}_{i_1 \cdots i_L}^{(L)}(k) \prod_{r=1}^L \hat{k}_{i_r} \right] \bar{\delta}_{\rm m}({\bf k}) , \\
  \delta_{\rm ani}^{\rm new}({\bf x}) &\equiv \int \frac{d^3 k}{(2\pi)^3} e^{i {\bf k} \cdot {\bf x}}
 \frac{1}{2} b_{\rm g}^{(2)} {\cal G}_{ij}^{(2)}(k) \hat{k}_i \hat{k}_j 
 \bar{\delta}_{\rm m}({\bf k}) , \label{eq:del}
 \end{split}
\end{align}
where $f$ is the linear growth rate, $a$ is the scale factor, $H$ is the Hubble parameter, and $\alpha \equiv d \ln n_{\rm av}(x) / d \ln x +  2$ is the selection function of a given galaxy sample. We note that $\bar{\delta}$ corresponds to the isotropy-conserving component, and $\delta_{\rm ani}^{\rm std}$ and $\delta_{\rm ani}^{\rm new}$ denote the isotropy-breaking ones depending on $b_{\rm g}$ and $b_{\rm g}^{(2)}$, respectively. Taking $f \to 0$ in eq.~\eqref{eq:del} yields the representation of $\delta^r$ up to linear order of ${\cal G}_{i_1 \cdots i_L}^{(L)}$.

In contrast, the velocity field is free from the above bias effect and therefore  reads
\begin{align}
\begin{split}
  u({\bf x}) &= \bar{u}({\bf x}) + u_{\rm ani}^{\rm std}({\bf x}) + u_{\rm ani}^{\rm new}({\bf x}), \\
  \bar{u}({\bf x}) &\equiv \int \frac{d^3 k}{(2\pi)^3} e^{i {\bf k} \cdot {\bf x}} \, i \frac{aHf}{k} (\hat{k} \cdot \hat{x}) \bar{\delta}_{\rm m} ({\bf k}) , \\
  u_{\rm ani}^{\rm std}({\bf x}) &\equiv \int \frac{d^3 k}{(2\pi)^3} e^{i {\bf k} \cdot {\bf x}} \,
  i \frac{aHf}{k} (\hat{k} \cdot \hat{x}) \left[ \frac{1}{2} \sum_{L > 0} {\cal G}_{i_1 \cdots i_L}^{(L)}(k) \prod_{r=1}^L \hat{k}_{i_r} \right] \bar{\delta}_{\rm m} ({\bf k}), \\
  u_{\rm ani}^{\rm new}({\bf x}) &\equiv 0 . \label{eq:u}
  \end{split}
\end{align}
Regarding the velocity field, up to linear order of $\bar{\delta}_{\rm m}$, the real-space expression coincides with this redshift-space one \cite{Okumura:2013zva}.

\subsection{Spin-2 galaxy field: ellipticity ${}_{\pm 2}\gamma$}

The ellipticity field, $\gamma_{ij}$, is defined as the transverse and traceless projection of the second moment of the surface brightness of galaxies $g_{ij}^I$ [see eq.~\eqref{eq:gij_calG}], namely, 
\begin{align}
  \gamma_{ij}({\bf x}) = \frac{1}{2} \left[ P_{ik}(\hat{x}) P_{j l}(\hat{x})
+ P_{il}(\hat{x}) P_{jk}(\hat{x})
- P_{ij}(\hat{x}) P_{k l}(\hat{x}) \right] g_{k l}^I ({\bf x})  ,
\end{align}
where $P_{ij}(\hat{x}) \equiv \delta_{ij}^{\rm K} - \hat{x}_i \hat{x}_j$. A conventionally-used $+$/$\times$ state is defined as 
\begin{align}
  \begin{pmatrix}
    \gamma_+ \\
    \gamma_\times 
  \end{pmatrix}
  ({\bf x})
  \equiv
  \begin{pmatrix}
    \hat{\theta}_i(\hat{x}) \hat{\theta}_{j}(\hat{x}) - \hat{\phi}_{i}(\hat{x}) \hat{\phi}_{j}(\hat{x}) \\
    \hat{\theta}_i(\hat{x}) \hat{\phi}_{j}(\hat{x}) + \hat{\phi}_{i}(\hat{x}) \hat{\theta}_{j}(\hat{x})
  \end{pmatrix}
   \gamma_{ij}({\bf x}) , \label{eq:gam_+x_def}
\end{align}
where $\hat{x} \equiv (\sin\theta \cos\phi, \sin\theta \sin\phi, \cos\theta)$, $\hat{\theta}(\hat{x}) \equiv (\cos\theta \cos\phi, \cos\theta \sin\phi, -\sin\theta)$ and $\hat{\phi}(\hat{x}) \equiv (- \sin\phi, \cos\phi, 0)$ are three orthonormal vectors. Here, for good compatibility with the later PolypoSH decomposition, we also introduce a helicity $\pm 2$ state as
\begin{align}
{}_{\pm 2}\gamma({\bf x}) \equiv m_{\mp}^i(\hat{x}) m_{\mp}^j(\hat{x}) \gamma_{ij}({\bf x}) ,  \label{eq:gam_lam_def}
\end{align}
where the polarization vector, given by
\begin{align}
 {\bf m}_{\pm}(\hat{x}) \equiv \frac{1}{\sqrt{2}}
  \left[ \hat{\theta}(\hat{x}) \mp i \hat{\phi}(\hat{x}) \right], \label{eq:pol_vec_def} 
\end{align}
obeys $\hat{x} \cdot {\bf m}_{\pm}(\hat{x}) = 0$, ${\bf m}_{\pm}^*(\hat{x}) = {\bf m}_{\mp}(\hat{x}) = {\bf m}_{\pm}(-\hat{x})$ and ${\bf m}_{\lambda}(\hat{x}) \cdot {\bf m}_{\lambda'}(\hat{x}) = \delta_{\lambda, -\lambda'}^{\rm K}$.
These two different states are linearly connected to each other and hence
\begin{align}
    {}_{\pm 2}\gamma({\bf x}) = \frac{1}{2} \left[ \gamma_+({\bf x}) \pm i \gamma_\times({\bf x}) \right]. \label{eq:gam_lam_vs_gam_+x}
  \end{align}

Similarly to the spin-0 case, ${\cal G}_{i_1 \cdots i_L}^{(L)}$ is likely to affect the bias relation between $\delta_{\rm m}$ and $g_{ij}^I$ (and hence ${}_{\pm 2}\gamma$). Fully expanding $g_{ij}^I$ in terms of $\frac{\partial_i \partial_j}{\partial^2} \delta_{\rm m}$ by use of $\delta_{ij}^{\rm K}$ and ${\cal G}_{i_1 \cdots i_L}^{(L)}$ for tensor contractions, we find the expression at linear order of $\delta_{\rm m}$:
\begin{align}
    g_{ij}^I({\bf x}) &= \int \frac{d^3 k}{(2\pi)^3} e^{i {\bf k} \cdot {\bf x}}
    \left[ b_{\rm K} \left( \hat{k}_i \hat{k}_j - \frac{1}{3} \delta_{ij}^{\rm K} \right)
      +  
      \frac{1}{2} b_{\rm K}^{(2,0)} {\cal G}_{ij}^{(2)}(k) 
       \right. \nonumber \\    
&\left. \quad\qquad\qquad\qquad + \frac{1}{4} b_{\rm K}^{(2,2)}
    \left( {\cal G}_{il}^{(2)}(k) \hat{k}_l \hat{k}_j
    + {\cal G}_{jl}^{(2)}(k) \hat{k}_l \hat{k}_i
    - \frac{2}{3} {\cal G}_{kl}^{(2)}(k) \hat{k}_k \hat{k}_l \delta_{ij}^{\rm K}
    - \frac{2}{3} {\cal G}_{ij}^{(2)}(k)
    \right)  \right. \nonumber \\    
&\left. \quad\qquad\qquad\qquad + \frac{1}{2} b_{\rm K}^{(4)} {\cal G}_{ijkl}^{(4)}(k) \hat{k}_k \hat{k}_l
    \right] \delta_{\rm m}({\bf k}) , \label{eq:gij_calG}
\end{align}
where $b_{\rm K}$ is equivalent to a linear bias parameter introduced in standard isotropy-conserving universe models, while the other three, $b_{\rm K}^{(2,0)}$, $b_{\rm K}^{(2,2)}$ and $b_{\rm K}^{(4)}$, are newly-introduced ones due to nonvanishing ${\cal G}_{ij}^{(2)}$ and ${\cal G}_{ijkl}^{(4)}$, or equivalently, nonvanishing $G_{2M}$ and $G_{4M}$. Converting this $g_{ij}^I$ into the spin-2 field ${}_{\pm 2}\gamma$ following the above conventions leads to the representation up to linear order of ${\cal G}_{i_1 \cdots i_L}^{(L)}$:
\begin{align}
\begin{split}
  {}_{\pm 2}\gamma({\bf x}) 
  &= {}_{\pm 2}\bar{\gamma}({\bf x})
  + {}_{\pm 2}\gamma_{\rm ani}^{\rm std}({\bf x})
  + {}_{\pm 2}\gamma_{\rm ani}^{\rm new}({\bf x}) , \\
  {}_{\pm 2}\bar{\gamma}({\bf x})
  &\equiv  \int \frac{d^3 k}{(2\pi)^3} e^{i {\bf k} \cdot {\bf x}}
  b_{\rm K} \hat{k}_i \hat{k}_j m_i^{\mp}(\hat{x}) m_j^{\mp}(\hat{x}) \bar{\delta}_{\rm m}({\bf k}) , \\
  {}_{\pm 2}\gamma_{\rm ani}^{\rm std}({\bf x}) 
&\equiv \int \frac{d^3 k}{(2\pi)^3} e^{i {\bf k} \cdot {\bf x}}
    b_{\rm K} \hat{k}_i \hat{k}_j m_i^{\mp}(\hat{x}) m_j^{\mp}(\hat{x}) 
    \left[\frac{1}{2} \sum_{L > 0} {\cal G}_{i_1 \cdots i_L}^{(L)}(k) \prod_{r=1}^L \hat{k}_{i_r}  \right]
    \bar{\delta}_{\rm m}({\bf k}) , \\
{}_{\pm 2}\gamma_{\rm ani}^{\rm new}({\bf x}) 
&\equiv \int \frac{d^3 k}{(2\pi)^3} e^{i {\bf k} \cdot {\bf x}}
\frac{1}{2}
\left[ b_{\rm K}^{(2,0)} {\cal G}_{ij}^{(2)}(k) 
    + b_{\rm K}^{(2,2)} {\cal G}_{il}^{(2)}(k) \left( \hat{k}_l \hat{k}_j - \frac{1}{3}\delta_{lj}^{\rm K} \right) \right. \\    
&\left. \qquad\qquad\qquad\qquad  
    + b_{\rm K}^{(4)} {\cal G}_{ijkl}^{(4)}(k) \hat{k}_k \hat{k}_l
    \right]  m_i^{\mp}(\hat{x}) m_j^{\mp}(\hat{x})
\bar{\delta}_{\rm m}({\bf k}) , \label{eq:gam}
  \end{split}
\end{align}
where ${}_{\pm 2}\bar{\gamma}$ denotes the isotropy-conserving component, and ${}_{\pm 2}\gamma_{\rm ani}^{\rm std}$ and ${}_{\pm 2}\gamma_{\rm ani}^{\rm new}$ are the isotropy-breaking ones depending on $b_{\rm K}$ and $(b_{\rm K}^{(2,0)}, b_{\rm K}^{(2,2)}, b_{\rm K}^{(4)})$, respectively. Also about the ellipticity field, there is no distinction between the real and redshift space expressions at linear order of $\bar{\delta}_{\rm m}$.

\subsection{Unified form of the galaxy field}

For later convenience, let us express the above three fields (density $\delta$, velocity $u$ and ellipticity ${}_{\pm 2}\gamma$) using the spin-weighted spherical harmonics. In eqs.~\eqref{eq:del}, \eqref{eq:u} and \eqref{eq:gam}, we convert ${\cal G}_{i_1 \cdots i_L}^{(L)}$ into $G_{LM}$ with eq.~\eqref{eq:GLM_2_calG}, expand all vectors $\hat{k}$, $\hat{x}$, and ${\bf m}_{\pm}(\hat{x})$, with the spin-weighted spherical harmonics using eq.~\eqref{eq:math_expand}, and simplify the products of the resultant spin-weighted spherical harmonics using the addition theorem~\eqref{eq:math_Ylm} and \eqref{eq:math_wigner}. This computation procedure has been traditionally executed for the CMB polyspectrum computations \cite{Shiraishi:2010kd}. The bottom line form is as follow;
\begin{align}
\begin{split}
  {}_{\lambda}X({\bf x}) &= {}_\lambda \bar{X}({\bf x}) + {}_\lambda X_{\rm ani}^{\rm std}({\bf x}) + {}_\lambda X_{\rm ani}^{\rm new}({\bf x}) ,\\
  {}_{\lambda}\bar{X}({\bf x}) &\equiv
  \int \frac{d^3 k}{(2\pi)^3} e^{i {\bf k} \cdot {\bf x}}
  \sum_j  
    \frac{4\pi c_{j}^{X}(k) }{2j+1} \sum_\mu  Y_{j \mu}(\hat{k}) {}_{-\lambda} Y_{j \mu}^*(\hat{x})
   \bar{\delta}_{\rm m}({\bf k})
  , \\
    {}_{\lambda}X_{\rm ani}^{\rm std}({\bf x}) &\equiv
  \int \frac{d^3 k}{(2\pi)^3} e^{i {\bf k} \cdot {\bf x}}
   \sum_j  
    \frac{4\pi c_{j}^{X}(k) }{2j+1} \sum_\mu  Y_{j \mu}(\hat{k}) {}_{-\lambda} Y_{j \mu}^*(\hat{x})
     \frac{1}{2} \sum_{L>0} \sum_M G_{LM}(k) Y_{LM}(\hat{k}) \bar{\delta}_{\rm m}({\bf k}) , \\
  {}_{\lambda}X_{\rm ani}^{\rm new}({\bf x}) 
&\equiv \int \frac{d^3 k}{(2\pi)^3} e^{i {\bf k} \cdot {\bf x}}
 \frac{1}{2} \sum_{L j j'} e_{Ljj'}^{X}
\sum_{M \mu \mu'}
    G_{L M}(k)
  \begin{pmatrix}
    L & j & j'\\
    M & \mu & \mu' 
  \end{pmatrix}
Y_{j \mu}^*(\hat{k}) {}_{-\lambda}Y_{j' \mu'}^*(\hat{x})
\bar{\delta}_{\rm m}({\bf k}) , \label{eq:X}
  \end{split}
\end{align}
where $X = \{\delta, u, \gamma \}$ and
\begin{align}
  \begin{split}
    c_j^{\delta}(k) &= \left(b_{\rm g} + \frac{1}{3} f \right) \delta_{j, 0}^{\rm K}
    - i \frac{\alpha f}{k x} \delta_{j, 1}^{\rm K}
    + \frac{2}{3} f \delta_{j,2}^{\rm K} , \\
    c_j^u(k) &= i \frac{aHf}{k} \delta_{j, 1}^{\rm K}, \\
    c_j^{\gamma}(k) &= \frac{\sqrt{6}}{3} b_{\rm K} \delta_{j, 2}^{\rm K} , \\
    e_{Ljj'}^{\delta} &= \sqrt{20\pi} b_{\rm g}^{(2)} \delta_{L,2}^{\rm K} \delta_{j,2}^{\rm K} \delta_{j',0}^{\rm K} , \\
    e_{Ljj'}^{u} &= 0 , \\
    e_{Ljj'}^{\gamma} &= 
    \sqrt{30\pi} b_{\rm K}^{(2,0)} \delta_{L,2}^{\rm K} \delta_{j,0}^{\rm K} \delta_{j', 2}^{\rm K}
    -\sqrt{\frac{7 \pi }{3}} b_{\rm K}^{(2,2)} \delta_{L,2}^{\rm K} \delta_{j,2}^{\rm K} \delta_{j', 2}^{\rm K}
    + 3 \sqrt{\frac{21 \pi }{5}} b_{\rm K}^{(4)} \delta_{L,4}^{\rm K} \delta_{j,2}^{\rm K} \delta_{j', 2}^{\rm K} . \label{eq:X_coeff}
  \end{split} 
\end{align}
The subscript $\lambda$ in ${}_{\lambda}X$ represents the spin/helicity dependence of each field; thus, $\lambda = 0$ for $X = \delta, u$, and $\lambda = \pm 2$ for $X = \gamma$. Regarding the spin-0 fields, for notational simplicity, we sometimes omit the subscript $0$ in ${}_{0}X$ as in eqs.~\eqref{eq:del} and \eqref{eq:u}.  For $\lambda = 0$, ${}_\lambda \bar{X}$ in eq.~\eqref{eq:X} recovers the usual Legendre expansion.

\section{Efficient computation methodology for general types of the isotropy-breaking galaxy correlations} \label{sec:PolypoSH}

In this section, we shall build a computation methodology for the galaxy 2PCFs sourced from a general type of the isotropy-breaking matter power spectrum \eqref{eq:Pm_GLM}.

\subsection{Isotropy-breaking galaxy correlations}

Up to linear order of $G_{L>0, M}$, the 2PCF, $\xi_{\lambda_1 \lambda_2}^{X_1 X_2}({\bf x}_{12}, \hat{x}_1, \hat{x}_2)
  \equiv \Braket{{}_{\lambda_1}X_1({\bf x}_1) {}_{\lambda_2}X_2({\bf x}_2)}$, is computed as
  \begin{align}
    \begin{split}
  \xi_{\lambda_1 \lambda_2}^{X_1 X_2}({\bf x}_{12}, \hat{x}_1, \hat{x}_2)
  &= 
  \xi_{\lambda_1 \lambda_2 \, \rm std}^{X_1 X_2}({\bf x}_{12}, \hat{x}_1, \hat{x}_2)
  + \xi_{\lambda_1 \lambda_2 \, \rm new}^{X_1 X_2}({\bf x}_{12}, \hat{x}_1, \hat{x}_2), \\
  \xi_{\lambda_1 \lambda_2 \, \rm std}^{X_1 X_2}({\bf x}_{12}, \hat{x}_1, \hat{x}_2)
  &\equiv  \Braket{{}_{\lambda_1}\bar{X}_1({\bf x}_1) {}_{\lambda_2}\bar{X}_2({\bf x}_2)} \\
  &\quad + \Braket{{}_{\lambda_1}\bar{X}_1({\bf x}_1) {}_{\lambda_2}X_{2 \, \rm ani}^{\rm std}({\bf x}_2)} + \Braket{{}_{\lambda_1}X_{1 \, \rm ani}^{\rm std}({\bf x}_1) {}_{\lambda_2}\bar{X}_2({\bf x}_2)} , \\
  \xi_{\lambda_1 \lambda_2 \, \rm new}^{X_1 X_2}({\bf x}_{12}, \hat{x}_1, \hat{x}_2)
  &\equiv \Braket{{}_{\lambda_1}\bar{X}_1({\bf x}_1) {}_{\lambda_2}X_{2 \, \rm ani}^{\rm new}({\bf x}_2)} + \Braket{{}_{\lambda_1}X_{1 \, \rm ani}^{\rm new}({\bf x}_1) {}_{\lambda_2}\bar{X}_2({\bf x}_2)} .
  \end{split}
  \end{align}
  Regarding the 2PCFs composed of the density or ellipticity field, $\xi_{\lambda_1 \lambda_2 \, \rm std}^{X_1 X_2}$ and $\xi_{\lambda_1 \lambda_2 \, \rm new}^{X_1 X_2}$ denote the components depending on the standard $(b_{\rm g}, b_{\rm K})$ and new $(b_{\rm g}^{(2)}, b_{\rm K}^{(2,0)}, b_{\rm K}^{(2,2)} ,b_{\rm K}^{(4)})$ bias parameters, respectively. Since $\bar{\delta}_{\rm m}$ respects the statistical homogeneity, the 2PCF takes the form
  \begin{align}
\xi_{\lambda_1 \lambda_2}^{X_1 X_2}({\bf x}_{12}, \hat{x}_1, \hat{x}_2)
= \int \frac{d^3 k}{(2\pi)^3} e^{i {\bf k} \cdot {\bf x}_{12}}
P_{\lambda_1 \lambda_2}^{X_1 X_2}({\bf k}, \hat{x}_1, \hat{x}_2), \label{eq:xi}
\end{align}
  where ${\bf x}_{12} \equiv {\bf x}_1 - {\bf x}_2$. Computing $\xi_{\lambda_1 \lambda_2 \, \rm std}^{X_1 X_2}$ and $\xi_{\lambda_1 \lambda_2 \, \rm new}^{X_1 X_2}$ by use of eq.~\eqref{eq:X} and the addition theorem \eqref{eq:math_Ylm} and \eqref{eq:math_wigner} leads to
\begin{align}
  \begin{split}
  P_{\lambda_1 \lambda_2}^{X_1 X_2}({\bf k}, \hat{x}_1, \hat{x}_2)
  &= P_{\lambda_1 \lambda_2 \, \rm std}^{X_1 X_2}({\bf k}, \hat{x}_1, \hat{x}_2)
  + P_{\lambda_1 \lambda_2 \, \rm new}^{X_1 X_2}({\bf k}, \hat{x}_1, \hat{x}_2) , \\
  P_{\lambda_1 \lambda_2 \, \rm std}^{X_1 X_2}({\bf k}, \hat{x}_1, \hat{x}_2)
  &\equiv \bar{P}_{\rm m}(k)
  \sum_{LM} G_{L M}(k) Y_{L M}(\hat{k}) 
  \sum_{J j_1 j_2}  
  \frac{(4\pi)^2 (-1)^{j_2} h_{J j_1 j_2}^{0~0~0}}{(2j_1+1)(2j_2+1)}  c_{j_1}^{X_1}(k) c_{j_2}^{X_2}(k)  \\
  &\quad \times 
  \sum_{\mu \mu_1 \mu_2}
   \begin{pmatrix}
    J & j_1 & j_2\\
    \mu & \mu_1 & \mu_2 
  \end{pmatrix}
 Y_{J \mu}^*(\hat{k})   {}_{-\lambda_1} Y_{j_1 \mu_1}^*(\hat{x}_1) {}_{-\lambda_2} Y_{j_2 \mu_2}^*(\hat{x}_2) 
  , \\
 P_{\lambda_1 \lambda_2 \, \rm new}^{X_1 X_2}({\bf k}, \hat{x}_1, \hat{x}_2)
  &\equiv  
  \frac{1}{2} \bar{P}_{\rm m}(k) \sum_{L j_1 j_2 j'} \sum_{M \mu_1 \mu_2 \mu'}
  G_{L M}(k) Y_{j_1 \mu_1}(\hat{k}) (-1)^{j_2} Y_{j_2 \mu_2}(\hat{k})  \\ 
  &\quad
 \times \left[ 
  \frac{4\pi c_{j_1}^{X_1}(k) }{2j_1+1} 
    e_{L j_2 j'}^{X_2}
  \begin{pmatrix}
    L & j_2 & j'\\
    M & -\mu_2 & \mu' 
  \end{pmatrix}
  (-1)^{\mu_2}
      {}_{-\lambda_1} Y_{j_1 \mu_1}^*(\hat{x}_1)
  {}_{-\lambda_2}Y_{j' \mu'}^*(\hat{x}_2) 
  \right.  \\
  &\qquad \left. + 
  \frac{4\pi c_{j_2}^{X_2}(k) }{2j_2+1}  e_{L j_1 j'}^{X_1}  
  \begin{pmatrix}
    L & j_1 & j'\\
    M & -\mu_1 & \mu' 
  \end{pmatrix}
  (-1)^{\mu_1}
  {}_{-\lambda_2}Y_{j_2 \mu_2}^*(\hat{x}_2)
  {}_{-\lambda_1}Y_{j' \mu'}^*(\hat{x}_1) 
  \right] , \label{eq:P}
  \end{split} 
\end{align}
where $h_{l_1 l_2 l_3}^{s_1 s_2 s_3} \equiv \sqrt{\frac{(2 l_1 + 1)(2 l_2 + 1)(2 l_3 + 1)}{ 4 \pi}} \begin{pmatrix}
  l_1 & l_2 & l_3 \\
  s_1 & s_2 & s_3 
\end{pmatrix}$.
We note that $P_{\lambda_1 \lambda_2}^{X_1 X_2}({\bf k}, \hat{x}_1, \hat{x}_2)$ is not the Fourier counterpart of $\xi_{\lambda_1 \lambda_2}^{X_1 X_2}({\bf x}_{12}, \hat{x}_1, \hat{x}_2)$.

The relation between $\xi_{\lambda_1 \lambda_2}^{X_1 X_2}$ and the angular correlation function defined on the celestial sphere $\Braket{{}_{\lambda_1}a_{\ell_1 m_1}^{X_1} {}_{\lambda_2}a_{\ell_2 m_2}^{X_2}}$ is summarized in appendix~\ref{appen:Cl}.

\subsection{Spin-weighted tripolar spherical harmonic decomposition approach for the exact analysis}

As confirmed in refs.~\cite{Shiraishi:2016wec,Sugiyama:2017ggb,Akitsu:2019avy,Shiraishi:2020pea}, the PolypoSH decomposition approach is an efficient and fast way to compute the isotropy-breaking 2PCFs of the spin-0 fields such as the density $\delta$ and the velocity $u$. We here generalize it to deal with the higher-spin fields as the ellipticity ${}_{\pm 2}\gamma$. Our generalized formulae recover the 2PCFs obtained in ref.~\cite{Shiraishi:2020vvj} at the isotropy-conserving limit.

Let us start from the computation without assuming the PP approximation. Since the 2PCF \eqref{eq:xi} is characterized by $\hat{x}_{12}$, $\hat{x}_{1}$ and $\hat{x}_{2}$, let us introduce a basis function of these three directions, i.e., the spin-weighted TripoSH:
\begin{align}
 {}_{\lambda_1 \lambda_2} {\cal X}_{\ell \ell_1\ell_2 \ell'}^{LM}(\hat{x}_{12},\hat{x}_1,\hat{x}_2)
  &\equiv \{ Y_{\ell}(\hat{x}_{12}) \otimes \{ {}_{\lambda_1} Y_{\ell_1}(\hat{x}_1) \otimes {}_{\lambda_2}Y_{\ell_2}(\hat{x}_2) \}_{\ell'} \}_{LM} \nonumber \\
  &= \sum_{m m_1 m_2 m' } {\cal C}_{\ell m \ell' m'}^{LM} {\cal C}_{\ell_1 m_1 \ell_2 m_2}^{\ell' m'} 
  Y_{\ell m}(\hat{x}_{12}) {}_{\lambda_1}Y_{\ell_1 m_1}(\hat{x}_1) {}_{\lambda_2}Y_{\ell_2 m_2}(\hat{x}_2)  , \label{eq:TripoSH_def}
\end{align}
and perform the following decomposition:
\begin{align}
  \xi_{\lambda_1 \lambda_2}^{X_1 X_2}({\bf x}_{12}, \hat{x}_1, \hat{x}_2)
  = \sum_{\ell\ell_1\ell_2 \ell' LM} {}_{\lambda_1 \lambda_2}\Xi_{\ell\ell_1\ell_2 \ell'}^{LM X_1 X_2}(x_{12}) 
{}_{\lambda_1 \lambda_2}{\cal X}_{\ell\ell_1\ell_2\ell'}^{LM}(\hat{x}_{12},\hat{x}_1,\hat{x}_2) , \label{eq:TripoSH_xi_def} 
\end{align}
where ${\cal C}_{l_1 m_1 l_2 m_2}^{l_3 m_3} \equiv (-1)^{l_1 - l_2 + m_3} \sqrt{2l_3 + 1} \begin{pmatrix} l_1 & l_2 & l_3 \\ m_1 & m_2 & -m_3 \end{pmatrix}$ is the Clebsch-Gordan coefficient.

To obtain the TripoSH coefficient ${}_{\lambda_1 \lambda_2}\Xi_{\ell\ell_1\ell_2 \ell'}^{LM X_1 X_2}$, we also decompose $P_{\lambda_1 \lambda_2}^{X_1 X_2}$ as
\begin{align}
  P_{\lambda_1 \lambda_2}^{X_1 X_2}({\bf k}, \hat{x}_1, \hat{x}_2)
  = \sum_{\ell\ell_1\ell_2 \ell' LM} {}_{\lambda_1 \lambda_2}\Pi_{\ell\ell_1\ell_2 \ell'}^{LM X_1 X_2}(k) 
  {}_{\lambda_1 \lambda_2}{\cal X}_{\ell\ell_1\ell_2\ell'}^{LM}(\hat{k},\hat{x}_1,\hat{x}_2) .
\end{align}
Substituting this into eq.~\eqref{eq:xi}, performing the $\hat{k}$ integral by use of eq.~\eqref{eq:math_Ylm}, and comparing the resultant $\xi_{\lambda_1 \lambda_2}^{X_1 X_2}$ with eq.~\eqref{eq:TripoSH_xi_def}, we derive the Hankel transformation rule:
\begin{align}
  {}_{\lambda_1 \lambda_2}\Xi_{\ell\ell_1\ell_2\ell'}^{LM X_1 X_2}(x_{12}) 
  = i^{\ell} \int_0^\infty \frac{k^2 dk}{2\pi^2}  j_{\ell}(k x_{12})
  {}_{\lambda_1 \lambda_2}\Pi_{\ell \ell_1 \ell_2 \ell'}^{LM X_1 X_2}(k) . \label{eq:hankel}
\end{align}
This is computed after obtaining ${}_{\lambda_1 \lambda_2}\Pi_{\ell \ell_1 \ell_2 \ell'}^{LM X_1 X_2}$ by use of
\begin{align}
{}_{\lambda_1 \lambda_2} \Pi_{\ell \ell_1 \ell_2 \ell'}^{LM X_1 X_2}(k) 
  =  \int d^2 \hat{k}  \int d^2 \hat{x}_1  \int d^2 \hat{x}_2 \,
  P_{\lambda_1 \lambda_2}^{X_1 X_2}({\bf k}, \hat{x}_1, \hat{x}_2)
  {}_{\lambda_1 \lambda_2}{\cal X}_{\ell \ell_1\ell_2 \ell'}^{LM *}(\hat{k},\hat{x}_1,\hat{x}_2) . \label{eq:TripoSH_Picoeff_def}
\end{align}

Let us estimate ${}_{\lambda_1 \lambda_2}\Pi_{\ell \ell_1 \ell_2 \ell'}^{LM X_1 X_2}$ from eq.~\eqref{eq:P}. Performing the spherical harmonic integrals and simplifying the resultant Wigner symbols by use of eqs.~\eqref{eq:math_Ylm} and \eqref{eq:math_wigner}, we obtain
\begin{align}
  \begin{split}
   {}_{\lambda_1 \lambda_2}\Pi_{\ell \ell_1 \ell_2 \ell'}^{LM X_1 X_2}(k)
  &= {}_{\lambda_1 \lambda_2}\Pi_{\ell \ell_1 \ell_2 \ell' \, \rm std}^{LM X_1 X_2}(k)
  + {}_{\lambda_1 \lambda_2}\Pi_{\ell \ell_1 \ell_2 \ell' \, \rm new}^{LM X_1 X_2}(k) , \\   
  {}_{\lambda_1 \lambda_2}\Pi_{\ell \ell_1 \ell_2 \ell' \, \rm std}^{LM X_1 X_2}(k) 
  &\equiv  \bar{P}_{\rm m}(k)  
G_{L M}(k)
\frac{(4\pi)^2 (-1)^{\lambda_1 + \lambda_2  + \ell_1 + L} h_{\ell_1 \ell_2 \ell'}^{0~0~0} h_{\ell  \ell'  L}^{000}}{(2\ell_1 + 1)(2\ell_2 + 1)\sqrt{2\ell' + 1} \sqrt{2L + 1}} 
c_{\ell_1}^{X_1} (k) c_{\ell_2}^{X_2} (k) 
 , \\
{}_{\lambda_1 \lambda_2}\Pi_{\ell \ell_1 \ell_2 \ell' \, \rm new}^{LM X_1 X_2}(k) 
   &\equiv  \frac{1}{2} \bar{P}_{\rm m}(k) 
  G_{L M}(k) (-1)^{\lambda_1 + \lambda_2}
  \sqrt{\frac{2\ell' + 1}{2L+1}}
 \sum_{j}  \\ 
  &\quad
  \times \left[
  \frac{4\pi c_{\ell_1}^{X_1}(k) }{2\ell_1 + 1} e_{L j \ell_2}^{X_2}
  h_{\ell_1 j \ell}^{000}
  \left\{ \begin{matrix}
   \ell &  \ell' & L \\
   \ell_2 &  j & \ell_1  
  \end{matrix} \right\}
  (-1)^{\ell + \ell_1 + \ell_2}
  \right.  \\
  &\qquad \left. + 
 \frac{4\pi c_{\ell_2}^{X_2}(k) }{2\ell_2 + 1} e_{L j \ell_1}^{X_1} 
  h_{\ell_2 j \ell}^{000}
  \left\{ \begin{matrix}
   \ell &  \ell' & L \\
   \ell_1 &  j & \ell_2  
  \end{matrix} \right\}
  (-1)^{\ell' }
  \right] .
  \end{split} \label{eq:TripoSH_Picoeff_GLM}
\end{align}
The number of nonvanishing multipoles is restricted by the selection rules in $c_{j}^{X}$, $e_{L j j'}^{X}$ [see eq.~\eqref{eq:X_coeff}] and $h_{l_1 l_2 l_3}^{s_1 s_2 s_3}$. From eq.~\eqref{eq:TripoSH_Picoeff_GLM}, it is apparent that nonvanishing $G_{LM}$ is linearly reflected on ${}_{\lambda_1 \lambda_2}\Pi_{\ell \ell_1 \ell_2 \ell'}^{L M X_1 X_2}$. Note that ${}_{\lambda_1 \lambda_2}\Pi_{\ell \ell_1 \ell_2 \ell'}^{00 X_1 X_2}$ or ${}_{\lambda_1 \lambda_2}\Xi_{\ell\ell_1\ell_2\ell'}^{00 X_1 X_2}$ completely recovers the TripoSH coefficient computed from the isotropy-conserving matter power spectrum in ref.~\cite{Shiraishi:2020vvj} because of $G_{00} = \sqrt{4\pi}$, $h_{\ell  \ell'  0}^{000} = (-1)^{\ell} \sqrt{\frac{2 \ell+1}{4\pi}} \delta_{\ell \ell'}^{\rm K}$ and ${}_{\lambda_1 \lambda_2}\Pi_{\ell \ell_1 \ell_2 \ell' \, \rm new}^{00 X_1 X_2} = 0$.

\begin{figure}[t]
      \begin{center}
        \includegraphics[width=1.\textwidth]{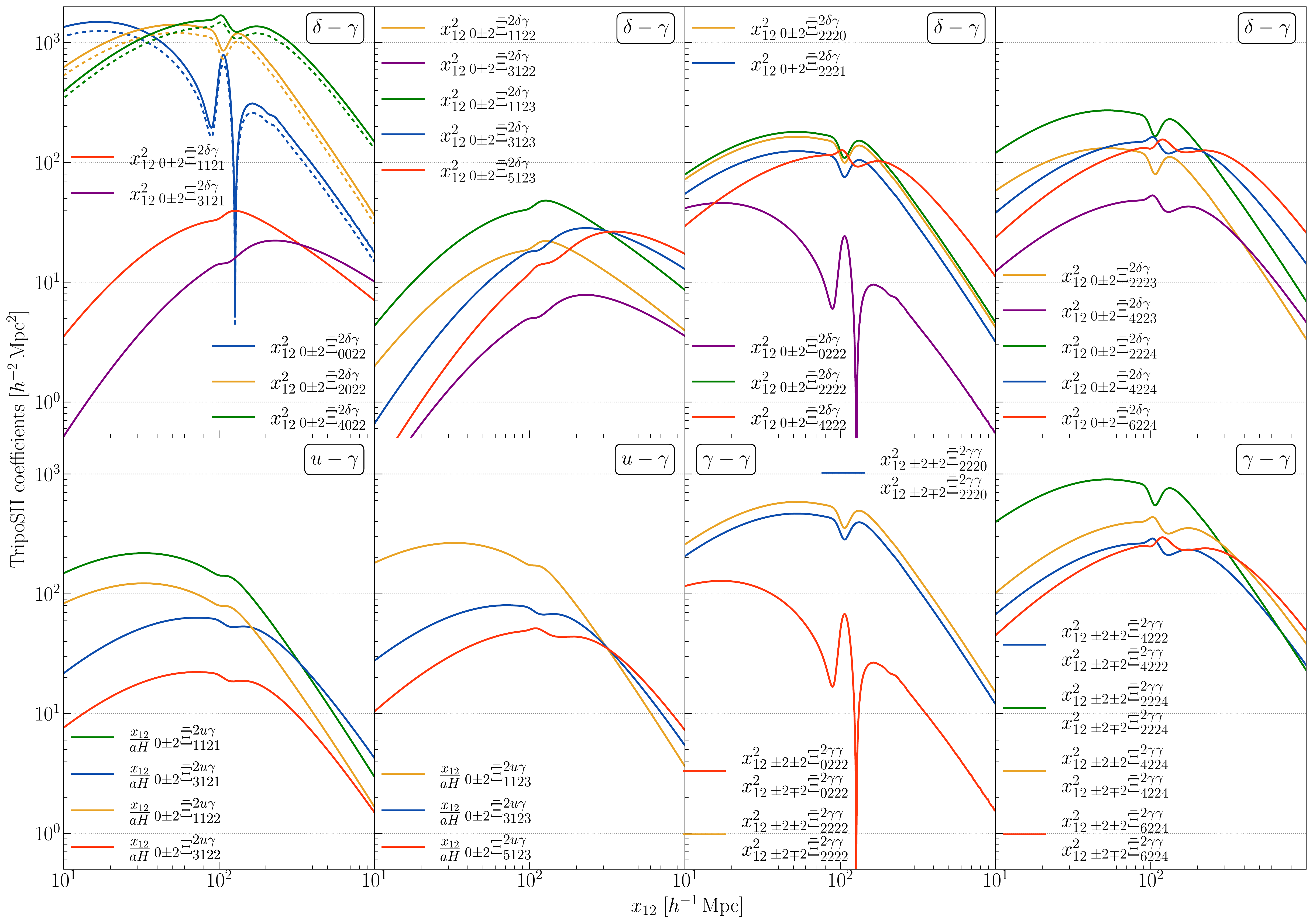}
      \end{center}
      \caption{Absolute values of all reduced TripoSH coefficients for $L = 2$: ${}_{0 \pm 2}\bar{\Xi}_{\ell\ell_1\ell_2\ell'}^{2 \delta \gamma}$ (top four panels), ${}_{0 \pm 2}\bar{\Xi}_{\ell\ell_1\ell_2\ell'}^{2 u \gamma}$ (bottom left two panels) and ${}_{\pm 2 \pm 2}\bar{\Xi}_{\ell\ell_1\ell_2\ell'}^{2 \gamma \gamma} = {}_{\pm 2 \mp 2}\bar{\Xi}_{\ell\ell_1\ell_2\ell'}^{2 \gamma \gamma}$ (bottom right two panels) as a function of $x_{12}$ for $z_1 = z_2 = 0.3$ and $b_{\rm g} = b_{\rm g}^{(2)} =  b_{\rm K} = b_{\rm K}^{(2,0)} = b_{\rm K}^{(2,2)} = 1$. Solid and dashed lines discriminate between the results in the redshift and real spaces, respectively. Note that the reduced TripoSH coefficient ${}_{\lambda_1 \lambda_2}\bar{\Xi}_{\ell\ell_1\ell_2\ell'}^{L X_1 X_2}$ is defined as the original TripoSH one ${}_{\lambda_1 \lambda_2}\Xi_{\ell\ell_1\ell_2\ell'}^{LM X_1 X_2}$ divided by $G_{LM}/\sqrt{4\pi}$ when $G_{LM} \propto k^0$ and therefore independent of $G_{LM}$.}
  \label{fig:xi_TripoSHcoeff_per_x12_L2}
\end{figure}

Figure~\ref{fig:xi_TripoSHcoeff_per_x12_L2} depicts all nonvanishing $L = 2$ TripoSH coefficients of the $\delta \gamma$, $u \gamma$ and $\gamma\gamma$ correlations as a function of $x_{12}$. Here we assume $G_{LM} \propto k^0$, so that $G_{LM}$ can be singled out in the TripoSH coefficient as ${}_{\lambda_1 \lambda_2}\Xi_{\ell\ell_1\ell_2\ell'}^{LM X_1 X_2} = \frac{G_{LM}}{ \sqrt{4\pi}} {}_{\lambda_1 \lambda_2}\bar{\Xi}_{\ell\ell_1\ell_2\ell'}^{L X_1 X_2} $. This figure rather shows the reduced coefficient ${}_{\lambda_1 \lambda_2}\bar{\Xi}_{\ell\ell_1\ell_2\ell'}^{2 X_1 X_2}$. At each multipole, the baryon acoustic oscillation (BAO) bump at $x_{12} \simeq 100 h^{-1} \, \rm Mpc$ is confirmed. It is also apparent that the redshift-space distortion (RSD) effect induces higher multipoles in the $\delta \gamma$ correlation, making the difference in shape between the real and redshift space 2PCFs as seen in section~\ref{sec:result} and appendix~\ref{appen:dv}.

The results in figure~\ref{fig:xi_TripoSHcoeff_per_x12_L2} are utilized for computing the 2PCFs of the $g_*$ model in section~\ref{sec:result}.

\subsection{Spin-weighted bipolar spherical harmonic decomposition approach for the plane-parallel-limit analysis}

In the following, we reduce the above expression to that in the PP limit. In this limit, the 2PCF is characterized by $\hat{x}_{12}$ and $\hat{x}_{\rm p}$ that is a LOS direction chosen as satisfying $\hat{x}_{\rm p} \simeq \hat{x}_1 \simeq \hat{x}_2$; therefore, the decomposition formula reduces to
\begin{align}
  \xi_{\lambda_1 \lambda_2  \rm PP}^{X_1 X_2}({\bf x}_{12}, \hat{x}_{\rm p})
  &\equiv  \xi_{\lambda_1 \lambda_2}^{X_1 X_2}({\bf x}_{12}, \hat{x}_{\rm p}, \hat{x}_{\rm p}) \nonumber \\
  &= \sum_{\ell \ell' LM} {}_{\lambda_1 \lambda_2}\xi_{\ell  \ell'}^{LM X_1 X_2}(x_{12}) 
     {}_{\lambda_1 + \lambda_2} X_{\ell \ell'}^{LM}(\hat{x}_{12},\hat{x}_{\rm p}) ,
  \label{eq:BipoSH_xi_def}
\end{align}
where the spin-weighted BipoSH basis is defined as
\begin{align}
 {}_{\lambda'} X_{\ell \ell'}^{LM}(\hat{x}_{12},\hat{x}_{\rm p})
 &\equiv \{Y_{\ell}(\hat{x}_{12}) \otimes {}_{\lambda'}Y_{\ell'}(\hat{x}_{\rm p})\}_{LM} \nonumber \\
 &= \sum_{mm'} {\cal C}_{\ell m \ell' m'}^{LM} Y_{\ell m}(\hat{x}_{12}) {}_{\lambda'}Y_{\ell' m'}(\hat{x}_{\rm p}). \label{eq:BipoSH_def}
\end{align}

In a similar manner to the TripoSH decomposition, we perform the Fourier-space predecomposition:
\begin{align}
  P_{\lambda_1 \lambda_2 \rm PP}^{X_1 X_2}({\bf k}, \hat{x}_{\rm p})
  &\equiv P_{\lambda_1 \lambda_2}^{X_1 X_2}({\bf k}, \hat{x}_{\rm p}, \hat{x}_{\rm p}) \nonumber  \\
  &= \sum_{\ell \ell' LM} {}_{\lambda_1 \lambda_2} \pi_{\ell \ell'}^{LM X_1 X_2}(k) 
  {}_{\lambda_1 + \lambda_2}X_{\ell \ell'}^{LM}(\hat{k},\hat{x}_{\rm p}).
\end{align}
With this, eqs.~\eqref{eq:xi}, \eqref{eq:BipoSH_xi_def} and the identities in appendix~\ref{appen:math}, we derive the Hankel transformation formula:
\begin{align}
  {}_{\lambda_1 \lambda_2}\xi_{\ell \ell'}^{LM X_1 X_2}(x_{12}) 
  = i^{\ell}  \int_0^\infty \frac{k^2 dk}{2\pi^2}
  j_{\ell}(k x_{12})  
  {}_{\lambda_1 \lambda_2} \pi_{\ell \ell'}^{LM X_1 X_2}(k) ,
  \label{eq:hankel_flat}
\end{align}
where
\begin{align}
  {}_{\lambda_1 \lambda_2}\pi_{\ell \ell'}^{LM X_1 X_2}(k) 
  =  \int d^2 \hat{k}  \int d^2 \hat{x}_{\rm p} \, 
  P_{\lambda_1 \lambda_2 \rm PP}^{X_1 X_2}({\bf k}, \hat{x}_{\rm p}) {}_{\lambda_1 + \lambda_2}X_{\ell \ell'}^{LM *}(\hat{k},\hat{x}_{\rm p}). \label{eq:BipoSH_picoeff_def}
  \end{align}

Because of the relation between the TripoSH and BipoSH bases:
\begin{align}
  {}_{\lambda_1 \lambda_2} {\cal X}_{\ell \ell_1\ell_2 \ell'}^{LM}(\hat{x}_{12},\hat{x}_{\rm p},\hat{x}_{\rm p})
 =  (-1)^{ \ell_1 - \ell_2 + \lambda_1 + \lambda_2 }
 \frac{h_{~ \ell_1 ~~~ \ell_2 ~~~ \ell'}^{-\lambda_1 ~ -\lambda_2 ~ \lambda_1 + \lambda_2}}{\sqrt{2\ell' + 1}}
      {}_{\lambda_1 + \lambda_2} X_{\ell \ell'}^{LM}(\hat{x}_{12},\hat{x}_{\rm p}) ,
\end{align}
the BipoSH coefficients can directly be inverted from the TripoSH ones as
\begin{align}
  \begin{split}
    {}_{\lambda_1 \lambda_2}\xi_{\ell  \ell'}^{LM X_1 X_2}(x_{12}) 
 &= \sum_{\ell_1\ell_2 } 
    (-1)^{ \ell_1 - \ell_2 + \lambda_1 + \lambda_2 }
    \frac{h_{~ \ell_1 ~~~ \ell_2 ~~~ \ell'}^{-\lambda_1 ~ -\lambda_2 ~ \lambda_1 + \lambda_2}}{\sqrt{2\ell' + 1}}
         {}_{\lambda_1 \lambda_2}\Xi_{\ell\ell_1\ell_2 \ell'}^{LM X_1 X_2}(x_{12}) , \\
      {}_{\lambda_1 \lambda_2}\pi_{\ell  \ell'}^{LM X_1 X_2}(k) 
      &=  \sum_{\ell_1\ell_2 } 
      (-1)^{ \ell_1 - \ell_2 + \lambda_1 + \lambda_2 }
      \frac{h_{~ \ell_1 ~~~ \ell_2 ~~~ \ell'}^{-\lambda_1 ~ -\lambda_2 ~ \lambda_1 + \lambda_2}}{\sqrt{2\ell' + 1}}
           {}_{\lambda_1 \lambda_2}\Pi_{\ell\ell_1\ell_2 \ell'}^{LM X_1 X_2}(k).
           \label{eq:TripoSH_Picoeff_vs_BipoSH_picoeff}
  \end{split}
\end{align}

Plugging eq.~\eqref{eq:P} into eq.~\eqref{eq:BipoSH_picoeff_def} and simplifying the integrals, or more simply, computing eq.~\eqref{eq:TripoSH_Picoeff_vs_BipoSH_picoeff} with eq.~\eqref{eq:TripoSH_Picoeff_GLM}, we obtain
\begin{align}
\begin{split}
  {}_{\lambda_1 \lambda_2}\pi_{\ell \ell'}^{LM X_1 X_2}(k)
  &= {}_{\lambda_1 \lambda_2}\pi_{\ell \ell' \, \rm std}^{LM X_1 X_2}(k) + {}_{\lambda_1 \lambda_2}\pi_{\ell \ell' \, \rm new}^{LM X_1 X_2}(k), \\
  {}_{\lambda_1 \lambda_2}\pi_{\ell \ell' \, \rm std}^{LM X_1 X_2}(k) 
  &\equiv \bar{P}_{\rm m}(k) G_{L M}(k) 
  \sum_{\ell_1 \ell_2} \frac{(4\pi)^2 (-1)^{\ell_2 + L}
h_{ ~ \ell_1 ~~~ \ell_2 ~~~ \ell'}^{ -\lambda_1 ~ -\lambda_2 ~ \lambda_1+\lambda_2}
h_{ \ell_1 \ell_2 \ell'}^{0~0~0}
h_{\ell \ell' L}^{000}}{(2\ell_1 + 1)(2\ell_2 + 1)(2\ell' + 1)\sqrt{2L + 1}}    
  c_{\ell_1}^{X_1} (k) c_{\ell_2}^{X_2} (k)
   , \\
  {}_{\lambda_1 \lambda_2}\pi_{\ell \ell' \, \rm new}^{LM X_1 X_2}(k) 
  &\equiv 
  \frac{1}{2} \bar{P}_{\rm m}(k) 
  G_{L M}(k) 
   \sum_{\ell_1 \ell_2}
   \frac{h_{~ \ell_1 ~~~ \ell_2 ~~~ \ell'}^{-\lambda_1 ~ -\lambda_2 ~ \lambda_1 +\lambda_2}}{\sqrt{2L+1}}
  \sum_{j}
    \\ 
  &\quad
 \times \left[ 
  \frac{4\pi c_{\ell_1}^{X_1}(k) }{2\ell_1+1} 
    e_{L j \ell_2}^{X_2}
    h_{\ell_1 j \ell}^{000} 
    \left\{
    \begin{matrix}
     \ell & \ell' & L \\
     \ell_2 & j & \ell_1
    \end{matrix}
    \right\}
    (-1)^{\ell}
  \right.  \\
  &\qquad \left. + 
  \frac{4\pi c_{\ell_2}^{X_2}(k) }{2\ell_2+1}  e_{L j \ell_1}^{X_1}  
  h_{\ell_2 j \ell}^{000} 
  \left\{
  \begin{matrix}
    \ell & \ell' & L \\
    \ell_1 & j & \ell_2
    \end{matrix}
  \right\}
 (-1)^{\ell_1 + \ell_2 + \ell'}
  \right] . \label{eq:BipoSH_Picoeff_GLM}
  \end{split}
\end{align}
The isotropy-conserving matter power spectrum induces only ${}_{\lambda_1 \lambda_2}\pi_{\ell \ell' }^{00 X_1 X_2}$ or ${}_{\lambda_1 \lambda_2}\xi_{\ell \ell'}^{00 X_1 X_2}$. The 2PCFs computed from these fully recover the results in ref.~\cite{Okumura:2019ned}.%
\footnote{The 2PCFs computed in ref.~\cite{Okumura:2019ned}, $\xi_{\rm g +}$, $\xi_{v +}$ and $\xi_{\pm}$, are related to ours according to $\xi_{\rm g +} =  \xi_{0 +2}^{\delta \gamma} +  \xi{}_{0 -2}^{\delta \gamma}$, $\xi_{v +} = \xi_{0 +2}^{u \gamma} + \xi_{0 -2}^{u \gamma}$ and $\xi_{\pm} = 2(\xi_{+2 \mp 2}^{\gamma\gamma} + \xi_{-2 \pm 2}^{\gamma\gamma} )$.}

\section{Correlations of the ellipticity field in the $g_*$ model} \label{sec:result}

\begin{figure}[t]
      \begin{center}
        \includegraphics[width=0.5\textwidth]{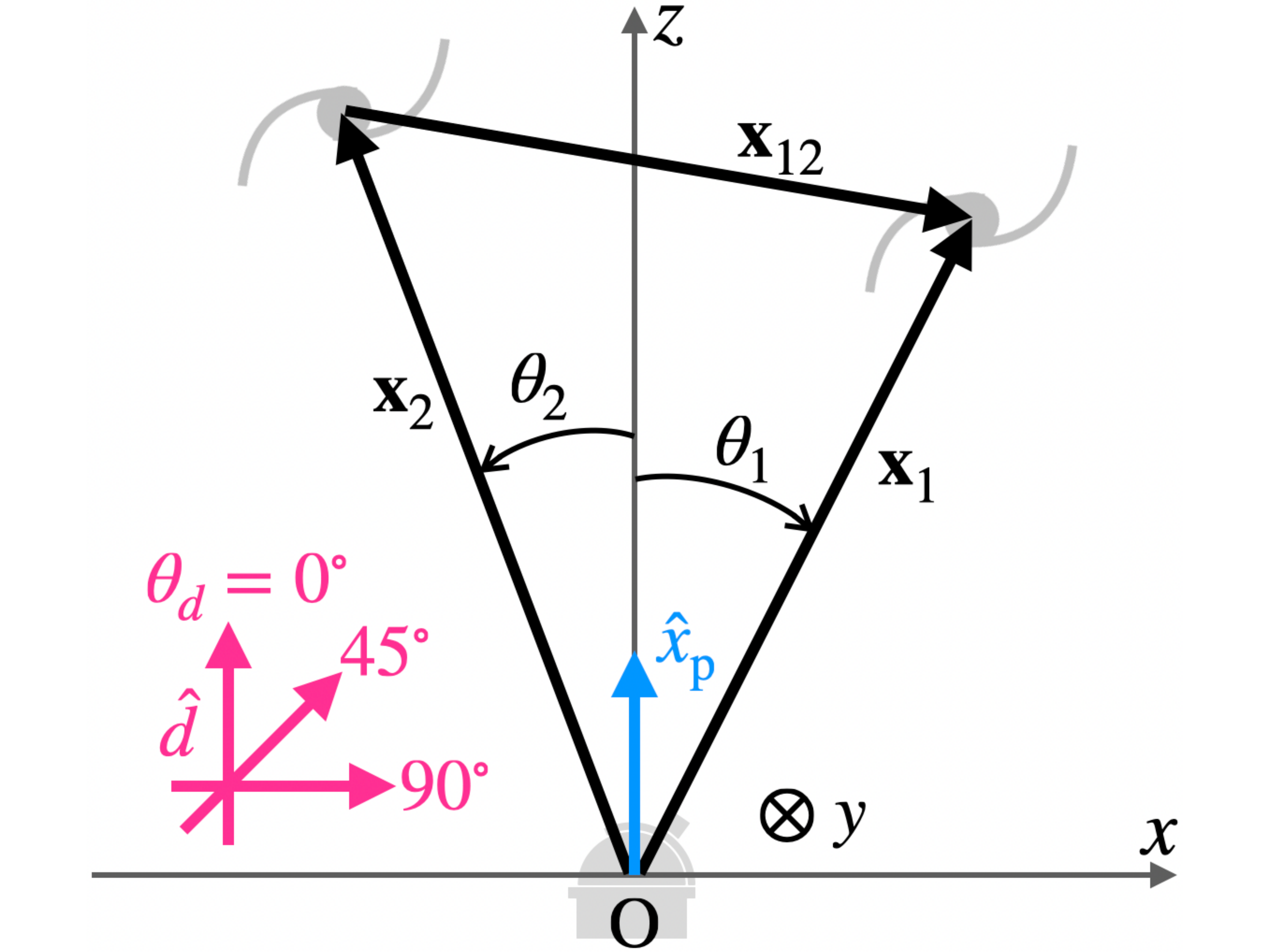}
      \end{center}
      \caption{Coordinate system adopted in the computation of the 2PCFs using the exact form $\xi_{\lambda_1 \lambda_2}^{X_1 X_2}({\bf x}_{12}, \hat{x}_1, \hat{x}_2)$ and the PP approximation $\xi_{\lambda_1 \lambda_2 \rm PP}^{X_1 X_2}({\bf x}_{12}, \hat{x}_{\rm p})$.}
  \label{fig:coordinate}
\end{figure}

Now we move to the numerical computation of the 2PCFs for a concrete scenario where the matter power spectrum includes a scale-invariant isotropy-breaking term whose magnitude is parametrized by $g_*$, reading 
\begin{align}
  P_{\rm m}({\bf k}) &= \bar{P}_{\rm m}(k)
  \left[ 1 + g_* \left\{ (\hat{k} \cdot \hat{d})^2 - \frac{1}{3} \right\} \right] \nonumber \\
  &= \bar{P}_{\rm m}(k)
  \left[ {\cal L}_0 (\hat{k} \cdot \hat{d}) + \frac{2}{3} g_*  {\cal L}_2 (\hat{k} \cdot \hat{d}) \right], \label{eq:Pm_gstar}
\end{align}
where $\hat{d}$ denotes some global preferred direction, and ${\cal L}_\ell(x)$ is the Legendre polynomial. With eq.~\eqref{eq:math_expand}, one can find the form of $G_{LM}$ in eq.~\eqref{eq:Pm_GLM} and ${\cal G}_{i_1 \cdots i_L}^{(L)}$ in eq.~\eqref{eq:Pm_calG} as
  \begin{align}
    \begin{split}
  G_{LM} &= \sqrt{4\pi} \delta_{L, 0}^{\rm K} \delta_{M, 0}^{\rm K}
  + \frac{8\pi}{15} g_* \, Y_{2M}^*(\hat{d}) \delta_{L, 2}^{\rm K} ,\\
  {\cal G}_{i_1 \cdots i_L}^{(L)} &= g_* \left( \hat{d}_{i_1} \hat{d}_{i_2} - \frac{1}{3} \delta_{i_1 i_2}^{\rm K} \right) \delta_{L,2}^{\rm K} .
  \end{split}
  \end{align}
  As confirmed in the previous section, the former ($G_{00}$) and latter ($G_{2M}$) terms make nonvanishing TripoSH coefficients ${}_{\lambda_1 \lambda_2}\Xi_{\ell\ell_1\ell_2\ell'}^{00 X_1 X_2}$ and ${}_{\lambda_1 \lambda_2}\Xi_{\ell\ell_1\ell_2\ell'}^{2M X_1 X_2}$, respectively. For late convenience, let us notationally differentiate the 2PCF to two terms as
\begin{align}
    \xi_{\lambda_1 \lambda_2}^{X_1 X_2}
  = {}^{(0)}\xi_{\lambda_1 \lambda_2}^{X_1 X_2}  
  + g_* \, {}^{(2)}\xi_{\lambda_1 \lambda_2}^{X_1 X_2}. 
  \label{eq:xi_L0_L2}
\end{align}
The former and latter terms correspond to the isotropy-conserving and isotropy-breaking parts, respectively.

The matter power spectrum \eqref{eq:Pm_GLM} for $G_{L>0, M} \neq 0$ can be generated by the existence of some anisotropic source. If there are (spin-1) vector fields which couple to inflaton or non-inflaton scalar ones in the inflationary era, nonvanishing $g_*$ or $G_{2M}$ arises as well as $G_{00}$ (e.g., refs.~\cite{Watanabe:2010fh,Bartolo:2012sd,Ohashi:2013qba,Bartolo:2014hwa,Naruko:2014bxa,Bartolo:2015dga,Abolhasani:2015cve,Fujita:2018zbr}). In more general, spin-$s$ fields produce nonzero $G_{00}$, $G_{2M}$, $G_{4M}$, $\cdots$, $G_{2(s-1), M}$ and $G_{2s, M}$ \cite{Kehagias:2017cym,Bartolo:2017sbu}. Other kinds of sources, e.g., two-form fields \cite{Obata:2018ilf}, an inflating solid or elastic medium \cite{Bartolo:2013msa, Bartolo:2014xfa}, fossil gravitational waves \cite{Masui:2010cz,Dai:2013kra,Schmidt:2013gwa,Akitsu:2022lkl} and large scale tides beyond the survey region \cite{Akitsu:2016leq,Akitsu:2017syq,Li:2017qgh,Chiang:2018mau,Akitsu:2019avy} also induce nonvanishing $G_{L > 0, M}$. The spectral shape of the induced $G_{LM}$ relies on the choice of, e.g., the coupling function and the potential of the fields. In the following 2PCF computations, let us focus on the simplest case where $g_*$ becomes constant in wavenumber. By the recent analysis with the CMB \cite{Planck:2018jri} and galaxy density fields \cite{Sugiyama:2017ggb}, $|g_*| > {\cal O}(10^{-2})$ is disfavored.

To make our discussion simpler, we shall work on the coordinate system where the three direction vectors $\hat{x}_{12}$, $\hat{x}_1$ and $\hat{x}_2$ are on the $xz$ plane and parametrized as $\hat{x}_{12} = (\sin \theta_{12}, 0, \cos\theta_{12})$, $\hat{x}_1 = (\sin \theta_1, 0, \cos\theta_1)$ and $\hat{x}_2 = (- \sin\theta_2, 0, \cos\theta_2)$ (here, the polar angles are fixed as $\phi_{12} = \phi_1 = 0$ and $\phi_2 = \pi$). In the PP limit; namely, $\theta_1 \to 0$ and $\theta_2 \to 0$, the two different LOS directions $\hat{x}_1$ and $\hat{x}_2$ can be identified with the $z$-axis positive direction, so that $\hat{x}_{\rm p} = (0, 0, 1)$. In the following, we examine how the 2PCFs are distorted depending on the preferred direction $\hat{d} = (\sin\theta_d \cos\phi_d, \sin\theta_d \sin\phi_d, \cos\theta_d)$ by considering three different cases: $\theta_d = 0^\circ$, $45^\circ$ and $90^\circ$ with $\phi_d = 0$, equivalently, $\hat{d} = (0, 0, 1)$, $(\frac{1}{\sqrt{2}}, 0, \frac{1}{\sqrt{2}})$ and $(1, 0, 0)$. For $\theta_d = 0^\circ$ and $90^\circ$, $\hat{d} = \hat{x}_{\rm p}$ and $\hat{d} \perp \hat{x}_{\rm p}$ hold, respectively. Even investigating $90^\circ \leq \theta_d \leq 180^\circ$, the symmetric results are obtained. Moreover, the distortion shapes depend weakly on the $y$-axis component of $\hat{d}$, so that we do not pick up any other $\hat{d}$. The coordinate system and settings mentioned above are visually summarized in figure~\ref{fig:coordinate}.


\begin{figure}[t]
  \begin{center}
    \includegraphics[width=0.93\textwidth]{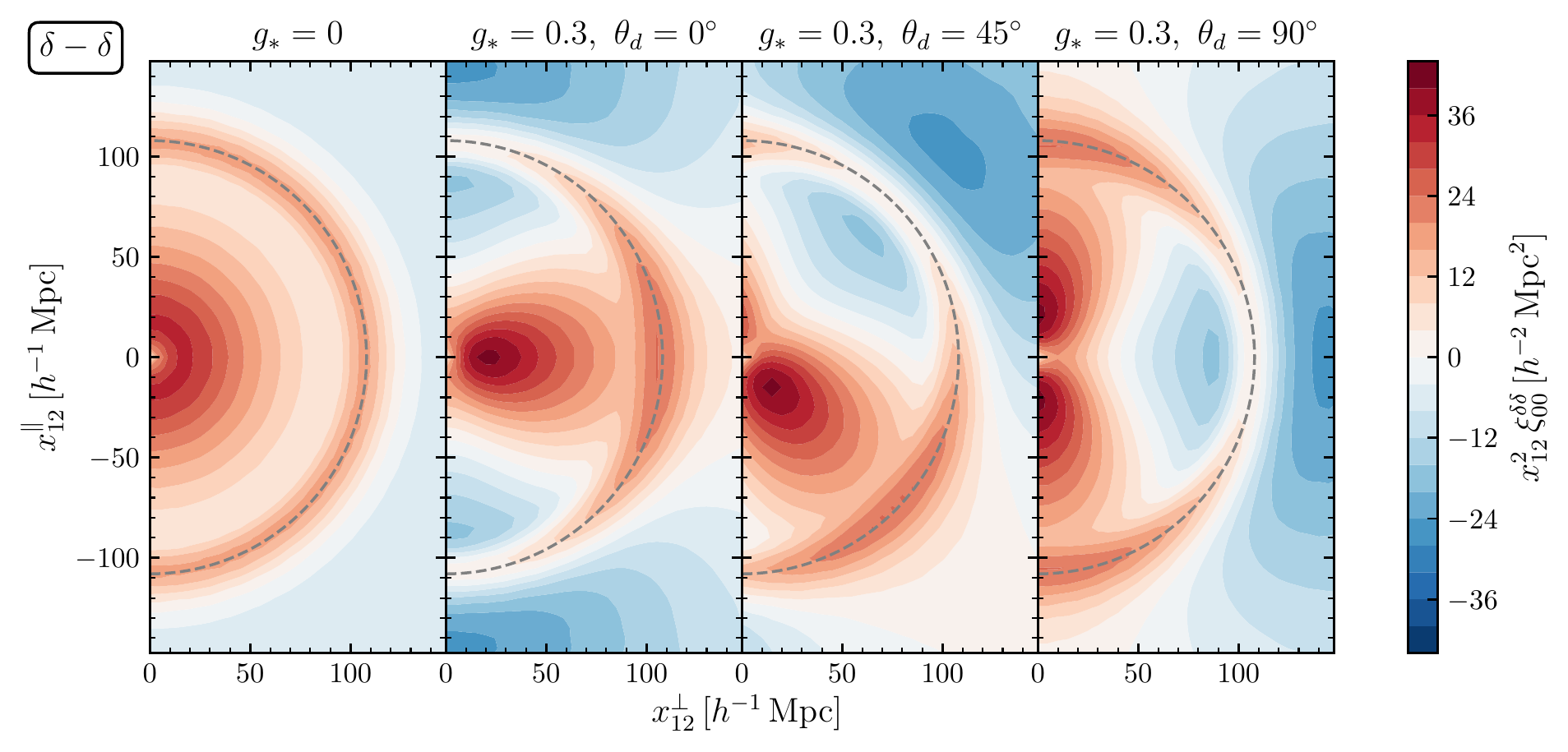} 
  \end{center}
  \caption{Intensity distributions of the real-space $\delta \delta$ correlation in the PP limit on the $(x_{12}^\perp, x_{12}^\parallel)$ plane for $z_1 = z_2 = 0.3$ and $b_{\rm g} = b_{\rm g}^{(2)} = 1$. The leftmost and the other three right panels show the results of the $g_*$ model for $g_* = 0$ and $0.3$, respectively (we consider such a large $g_*$ to highlight the isotropy-breaking signatures although it is disfavored by observations). For the latter, three different $\hat{d}$: $\theta_d = 0^\circ$, $45^\circ$ and $90^\circ$ are considered. The region for $x_{12}^\perp \geq 0$ is only shown, while one can reconstruct the region for $x_{12}^\perp < 0$ as these 2PCFs are symmetric or antisymmetric with respect to the origin. For reference, the BAO scale, $x_{12} \simeq 100 h^{-1} \, \rm Mpc$, is displayed with the dashed gray circles.}
  \label{fig:xi_flat_2D_dd}
\end{figure}

\begin{figure}[t]
  \begin{minipage}{1.\hsize}
    \begin{center}
      \includegraphics[width=0.93\textwidth]{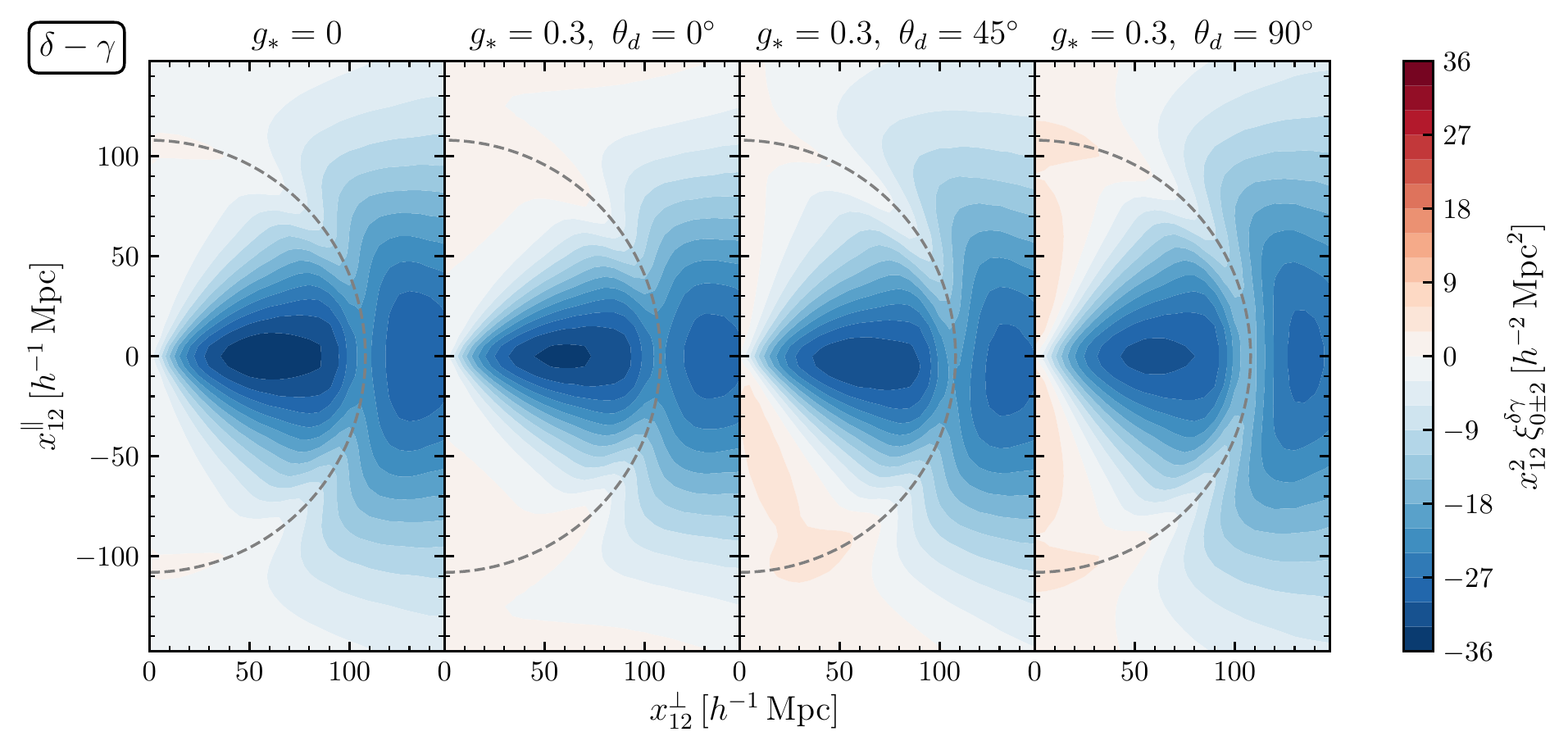}
    \end{center}
  \end{minipage}
  \\
  \begin{minipage}{1.\hsize}
    \begin{center}
      \includegraphics[width=0.93\textwidth]{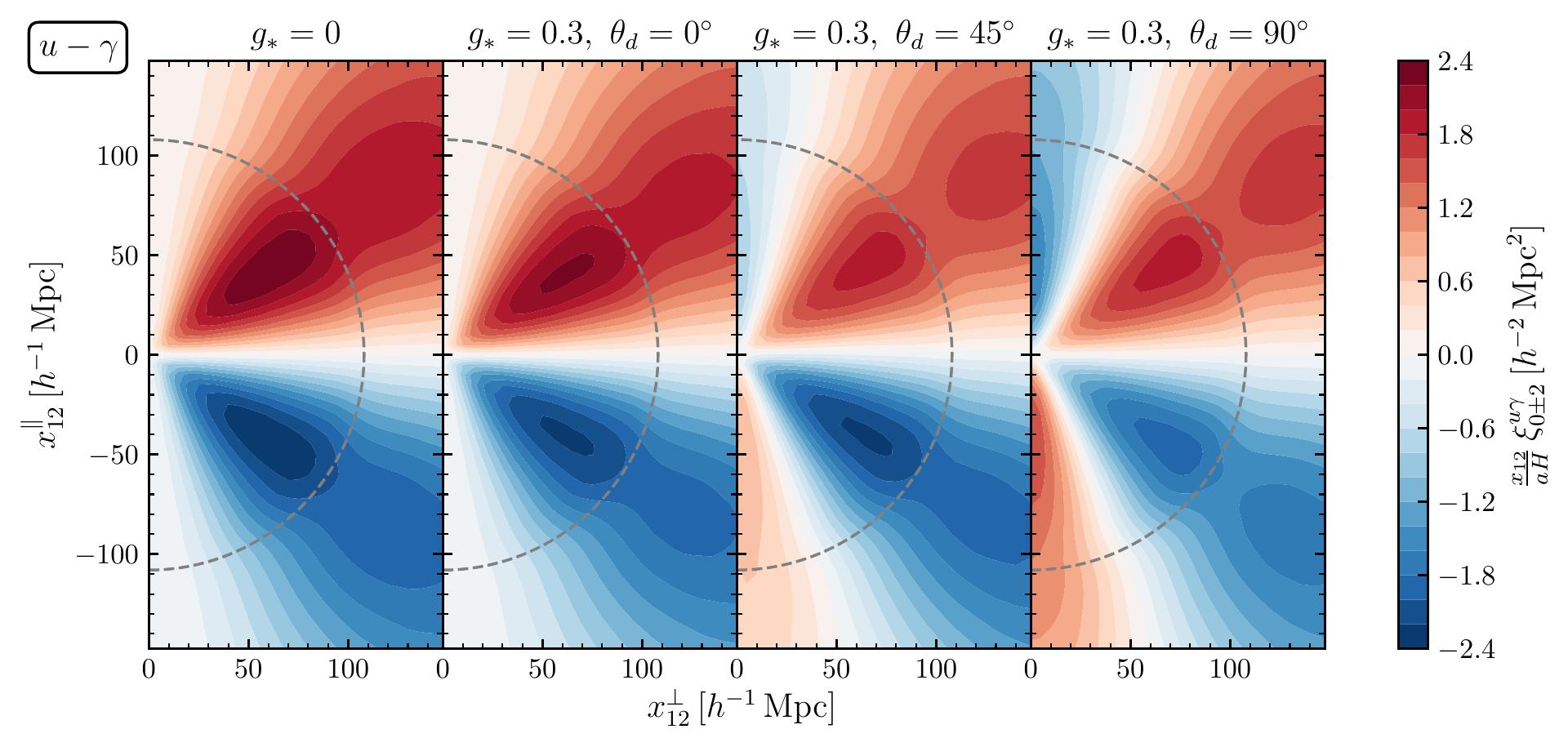}
    \end{center}
  \end{minipage}
  \caption{Same as figure~\ref{fig:xi_flat_2D_dd}, except for the $\delta \gamma$ correlation in the redshift space and the $u \gamma$ correlation when $b_{\rm g} = b_{\rm g}^{(2)} = b_{\rm K} = b_{\rm K}^{(2,0)} = b_{\rm K}^{(2,2)} = 1$.}
  \label{fig:xi_flat_2D_dr_ur}
\end{figure}

\begin{figure}[t]
  \begin{minipage}{1.\hsize}
    \begin{center}
      \includegraphics[width=0.93\textwidth]{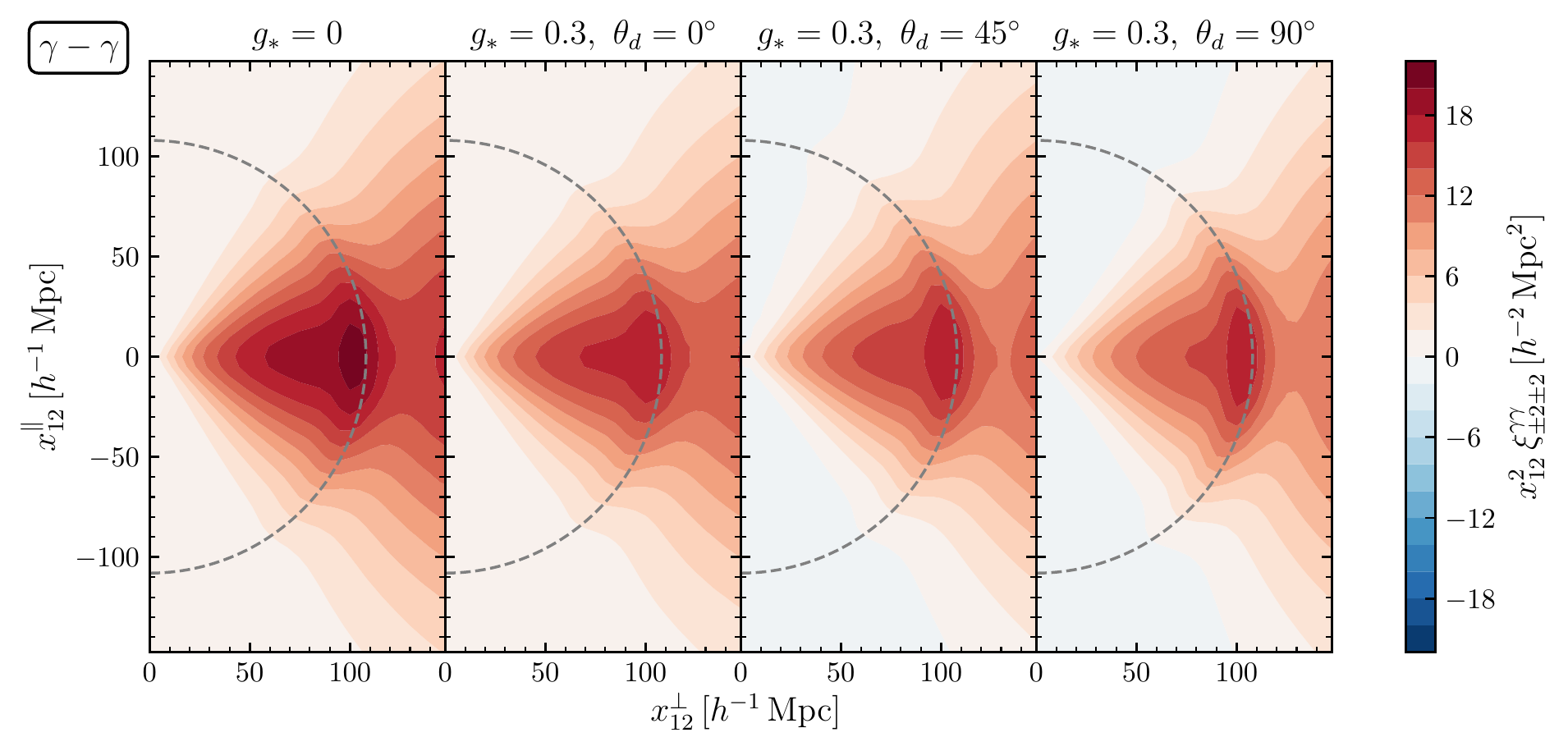}
    \end{center}
  \end{minipage} 
  \\
  \begin{minipage}{1.\hsize}
    \begin{center}
      \includegraphics[width=0.93\textwidth]{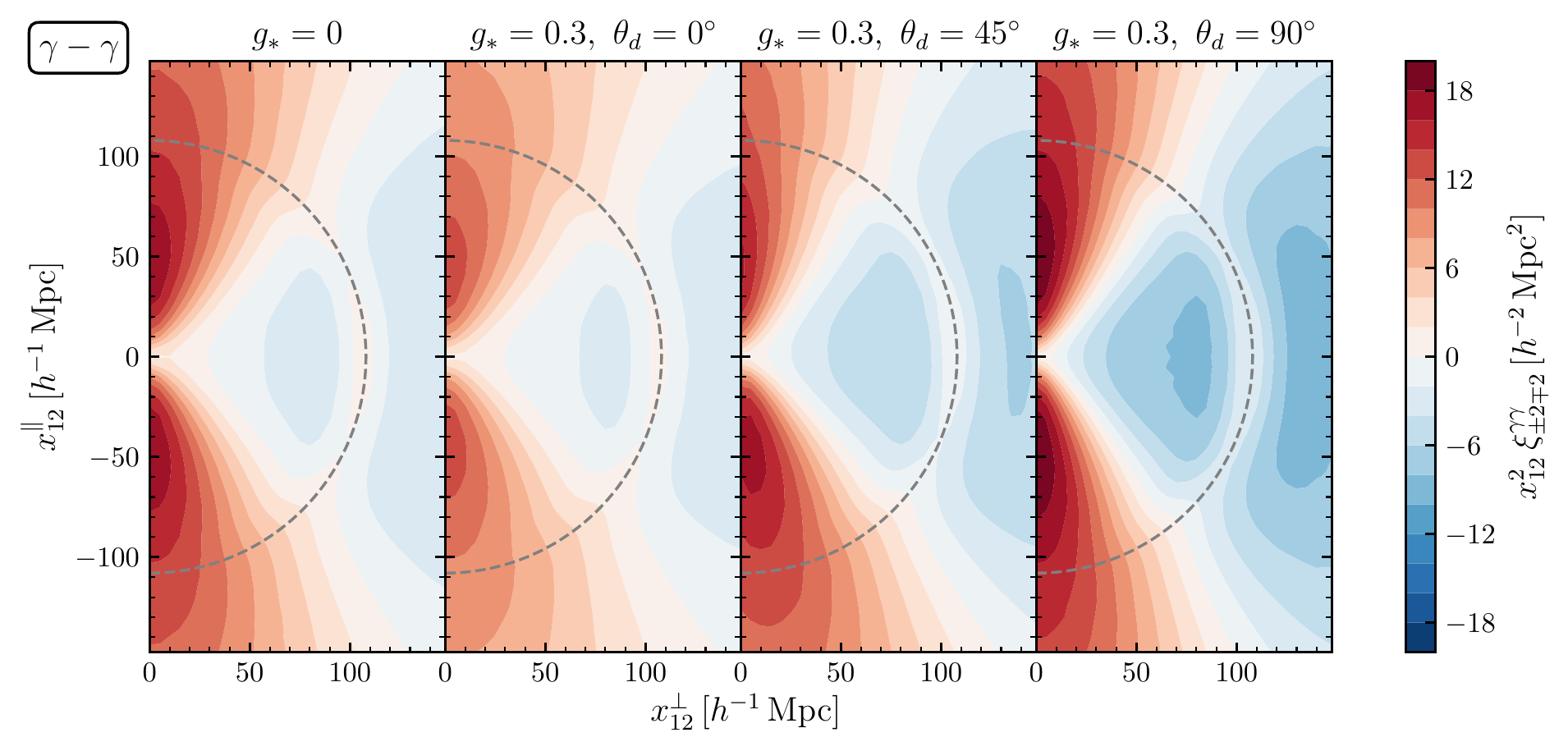}
    \end{center}
  \end{minipage}
  \caption{Same as figure~\ref{fig:xi_flat_2D_dd}, except for the $\gamma \gamma$ correlation when $b_{\rm K} = b_{\rm K}^{(2,0)} = b_{\rm K}^{(2,2)} = 1$.}
  \label{fig:xi_flat_2D_rr}
\end{figure}


We shall first assess the shape change of the 2PCFs depending on $\hat{d}$ based on the PP-limit results. For this purpose, we analyze $\xi_{\lambda_1 \lambda_2 \rm PP}^{X_1 X_2}({\bf x}_{12}, \hat{x}_{\rm p})$ as a function of $x_{12}^\perp$ and $x_{12}^\parallel$, the components of ${\bf x}_{12}$ perpendicular and parallel to the LOS direction $\hat{x}_{\rm p}$. In the computation, we set ${\bf x}_{12} = (x_{12}^\perp, 0, x_{12}^\parallel)$ and thus adopt $\theta_{12} = \arccos(\hat{x}_{12} \cdot \hat{x}_{\rm p}) = \arccos(x_{12}^\parallel / x_{12})$ and $x_{12} = \sqrt{(x_{12}^\parallel)^2 + (x_{12}^\perp)^2}$. Figures~\ref{fig:xi_flat_2D_dd}, \ref{fig:xi_flat_2D_dr_ur}, \ref{fig:xi_flat_2D_rr} and \ref{fig:xi_flat_2D_dd_dv_vv} depict the results for several $g_*$ and $\theta_d$.

Here, let us begin with the assessment of the $\delta \delta$ correlations in the real space as the isotropy-breaking signatures are most clearly apparent there. From figure~\ref{fig:xi_flat_2D_dd}, one can easily find that nonzero $g_*$ distorts the 2PCF depending on $\theta_d$. For $\theta_d = 0^\circ$, since $\hat{x}_{\rm p} = \hat{d}$, the 2PCF is distorted along the $x_{12}^\perp$ axis. This looks similar to the RSD effect in the redshift-space $\delta \delta$ correlation (top leftmost panel in figure~\ref{fig:xi_flat_2D_dd_dv_vv}). On the other hand, for $\theta_d = 90^\circ$ and $45^\circ$, the distortions seem to appear in a direction along the $x_{12}^\parallel$ axis and the diagonal line, respectively. Such a feature could be a distinctive indicator for testing the cosmic isotropy. For the other types of the 2PCFs described in figures~\ref{fig:xi_flat_2D_dr_ur}, \ref{fig:xi_flat_2D_rr} and \ref{fig:xi_flat_2D_dd_dv_vv}, the isotropy-breaking signatures due to nonzero $g_*$ are sometimes degenerate with the isotropy-conserving ones when $g_* = 0$, and the above trend is then slightly perceptible.

\begin{figure}[t]
      \begin{center}
        \includegraphics[width=1.\textwidth]{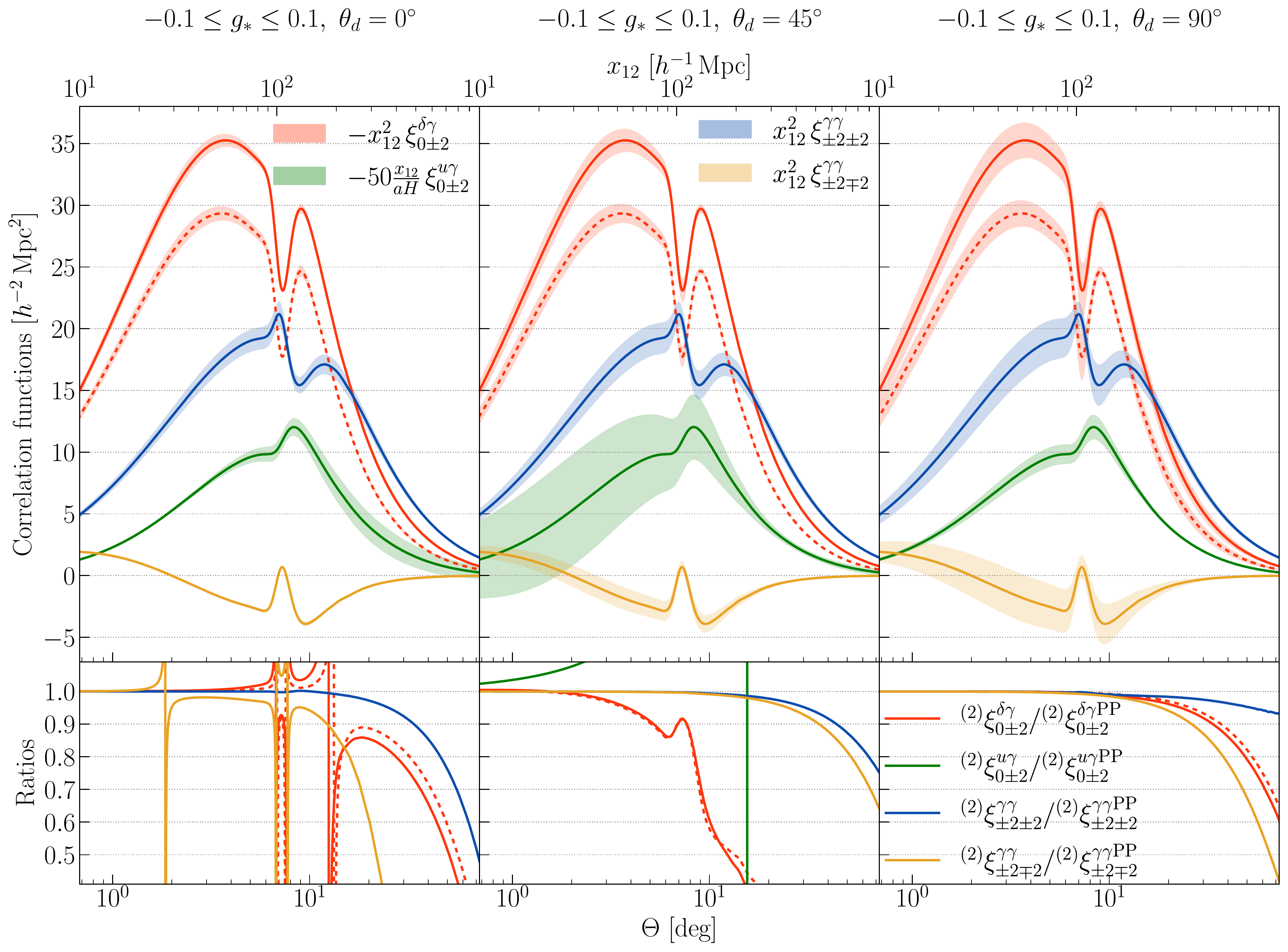}
      \end{center}
      \caption{Absolute values of the 2PCFs $\xi_{\lambda_1 \lambda_2}^{X_1 X_2}$ of the $g_*$ model for $-0.1 \leq g_* \leq 0.1$ (top panels) and ratios between the exact and PP-limit results of the isotropy-breaking parts ${}^{(2)}\xi_{\lambda_1 \lambda_2}^{X_1 X_2}$ (bottom panels) for the $\delta \gamma$, $u \gamma$ and $\gamma\gamma$ cases as a function of $\Theta$ or $x_{12}$ when $z_1 = z_2 = 0.3$ and $b_{\rm g} = b_{\rm g}^{(2)} =  b_{\rm K} = b_{\rm K}^{(2,0)} = b_{\rm K}^{(2,2)} = 1$. Here, three different $\hat{d}$: $\theta_d = 0^\circ$ (left panels), $45^\circ$ (center panels) and $90^\circ$ (right panels) are considered. In the top panels, the 2PCFs move under a range of $g_*$ and are thus expressed with the finite-width lines (the central lines correspond to the results for $g_* = 0$). Solid and dashed lines discriminate between the results in the redshift and real spaces, respectively. Note that all the results in the bottom panels are independent of $g_*$. In the bottom panels, some spiky features appear or some lines entirely disappear as ${}^{(2)}\xi_{\lambda_1 \lambda_2 \rm PP}^{X_1 X_2}$ sometimes vanish.}
  \label{fig:xi_per_Theta_dr_vr_rr}
\end{figure}

We further go through the 2PCF beyond the PP limit by use of the TripoSH decomposition technique. For numerical computations of $\xi_{\lambda_1 \lambda_2}^{X_1 X_2}({\bf x}_{12}, \hat{x}_1, \hat{x}_2)$,  we impose $\hat{x}_{12} = (1,0,0)$ (i.e., $\theta_{12} = 90^\circ$) and $\theta_1 = \theta_2$ and define the opening angle between $\hat{x}_1$ and $\hat{x}_2$ as $\Theta \equiv \theta_1 + \theta_2 = 2 \theta_1 = 2 \theta_2$. Due to these conditions, a triangle formed by ${\bf x}_{12}$, ${\bf x}_1$ and ${\bf x}_2$ becomes isosceles.

The results of the $\delta \gamma$, $u \gamma$ and $\gamma \gamma$ correlations including nonzero $g_*$ at several $x_{12}$ or $\Theta$ are summarized in figure~\ref{fig:xi_per_Theta_dr_vr_rr}. One can confirm from this that the distortion level of the 2PCF for a certain $g_*$ changes depending on $\theta_d$. It is enhanced as $\theta_d$ increases for the $\delta \gamma$ or $\gamma \gamma$ case, while maximized at $\theta_d \sim 45^\circ$ for the $u\gamma$ one (see the top panels). The latter feature is confirmed also from the $\delta u$ correlations (see figure~\ref{fig:xi_per_Theta_dd_dv_vv}).

From figure~\ref{fig:xi_per_Theta_dr_vr_rr}, one can also study how accurate the PP approximation is. The ratios between the exact and approximate results of the isotropy-breaking parts of the 2PCFs given in eq.~\eqref{eq:xi_L0_L2}, ${}^{(2)}\xi_{\lambda_1 \lambda_2}^{X_1 X_2} / {}^{(2)}\xi_{\lambda_1 \lambda_2 \rm PP}^{X_1 X_2}$, are plotted in the bottom panels. Basically, as $\Theta$ increases, this ratio departs from unity, meaning that the error of the PP approximation becomes larger. In the $\gamma \gamma$ case, for any $\theta_d$, the PP approximation works, at least, up to $\Theta \sim 10^\circ$ because the error is within $10\%$. Similar but better results are obtained in the $\delta \delta$ and $uu$ cases (see figure~\ref{fig:xi_per_Theta_dd_dv_vv}). In contrast, the $\delta \gamma$ case is sensitive to $\theta_d$, and if $\theta_d = 45^\circ$, the accuracy drops drastically and the error exceeds $10\%$ already at $\Theta \sim  5^\circ$. For the $u \gamma$ case, even worse, the PP approximation is totally spoiled. These indicate the importance of the analysis without the PP approximation for an accurate estimation of $g_*$. We note that, regarding the isotropy-conserving part ${}^{(0)}\xi_{\lambda_1 \lambda_2}^{X_1 X_2}$, for the $\delta \gamma$ and $\gamma \gamma$ cases, the PP approximation works under $\Theta \lesssim 30^\circ$ as shown in ref.~\cite{Shiraishi:2020vvj}.

\section{Conclusions} \label{sec:conclusion}

In this paper, we have studied the impacts of the cosmic isotropy breaking on the 2PCFs of the galaxy intrinsic alignment or equivalently the spin-2 galaxy ellipticity field for the first time. To achieve this, we first have developed a formalism for general types of the isotropy-breaking 2PCFs and an efficient computation methodology of them by generalizing the standard PolypoSH decomposition to the spin-weighted version. This is applicable to not only the analysis with the PP approximation but also the exact analysis. The previous spin-0 version of our computation methodology has already been implemented also for measuring galaxy clustering \cite{Sugiyama:2017ggb}. Since the new spin-weighted one is comparably speedy and efficient, it could also work in the data analysis with the ellipticity field.

As a concrete demonstration, we have analyzed the 2PCFs in a well-known $g_*$ model according to this methodology. It has been confirmed that some isotropy-breaking distortions appear in the 2PCFs, and their shapes rely on a preferred direction causing the isotropy violation $\hat{d}$. Such a feature could be a distinctive indicator for testing the cosmic isotropy.%
\footnote{
There remain some parameter degeneracies between the intrinsic isotropy-breaking parameters $G_{LM}$ and the galaxy bias parameters in the 2PCFs (see sections~\ref{sec:linear} and \ref{sec:PolypoSH}); hence, a careful treatment is necessary in the practical data analysis.
}
Comparing between the exact and the PP-limit results, we have quantified the error of the PP approximation as a function of a opening angle between the LOS directions towards target galaxies $\Theta$. For the ellipticity auto correlation, the error does not exceed $10\%$ at $\Theta \lesssim 10^\circ$ for any $\hat{d}$. For the density-ellipticity and velocity-ellipticity cross correlations, the error is enhanced for specific $\hat{d}$, and the validity of the PP approximation is no longer guaranteed even at $\Theta = {\cal O}(1^\circ)$. This suggests the importance of the analysis beyond the PP approximation for an accurate isotropy test. 

In the practical analysis, we have focused on the $g_*$ model, in which the matter power spectrum is given by eq.~\eqref{eq:Pm_gstar}, predicting nonzero TripoSH monopole and quadrupole, ${}_{\lambda_1 \lambda_2}\Xi_{\ell\ell_1\ell_2 \ell'}^{00 X_1 X_2}$ and ${}_{\lambda_1 \lambda_2}\Xi_{\ell\ell_1\ell_2 \ell'}^{2 M X_1 X_2}$. On the other hand, in other models that could yield nonzero higher multipoles ${}_{\lambda_1 \lambda_2}\Xi_{\ell\ell_1\ell_2 \ell'}^{L > 2, M X_1 X_2}$ as mentioned in section~\ref{sec:result}, the 2PCFs would be distorted in a different fashion. Moreover, including contributions due to not only the matter distribution (scalar mode) but also the vorticity (vector mode) and the gravitational wave (tensor mode) \cite{Schmidt:2012ne,Biagetti:2020lpx,Akitsu:2022lkl} might yield other unique shapes in the 2PCFs. These could also be dealt with by our spin-weighted PolypoSH decomposition methodology owing to its high versatility, and should be studied in future works.

Our results have been obtained on the basis of the linear theory and hence might increase ambiguity at smaller scales. It should be checked with N-body simulations and the higher-order perturbation theory as done in ref.~\cite{Okumura:2020hhr}.

\acknowledgments

M.\,S. is supported by JSPS KAKENHI Grant Nos.~JP19K14718, JP20H05859 and JP23K03390. T.\,O. acknowledges support from the Ministry of Science and Technology of Taiwan under Grant Nos. MOST 110-2112-M-001-045- and MOST 111-2112-M-001-061- and the Career Development Award, Academia Sinica (AS-CDA-108-M02) for the period of 2019 to 2023. K.\,A. is supported by JSPS Overseas Research Fellowships. M.\,S. and K.\,A. also acknowledge the Center for Computational Astrophysics, National Astronomical Observatory of Japan, for providing the computing resources of the Cray XC50. 



\appendix

\section{Correlations of the density and velocity fields in the $g_*$ model}\label{appen:dv}

\begin{figure}[t]
  \begin{minipage}{1.\hsize}
    \begin{center}
      \includegraphics[width=0.93\textwidth]{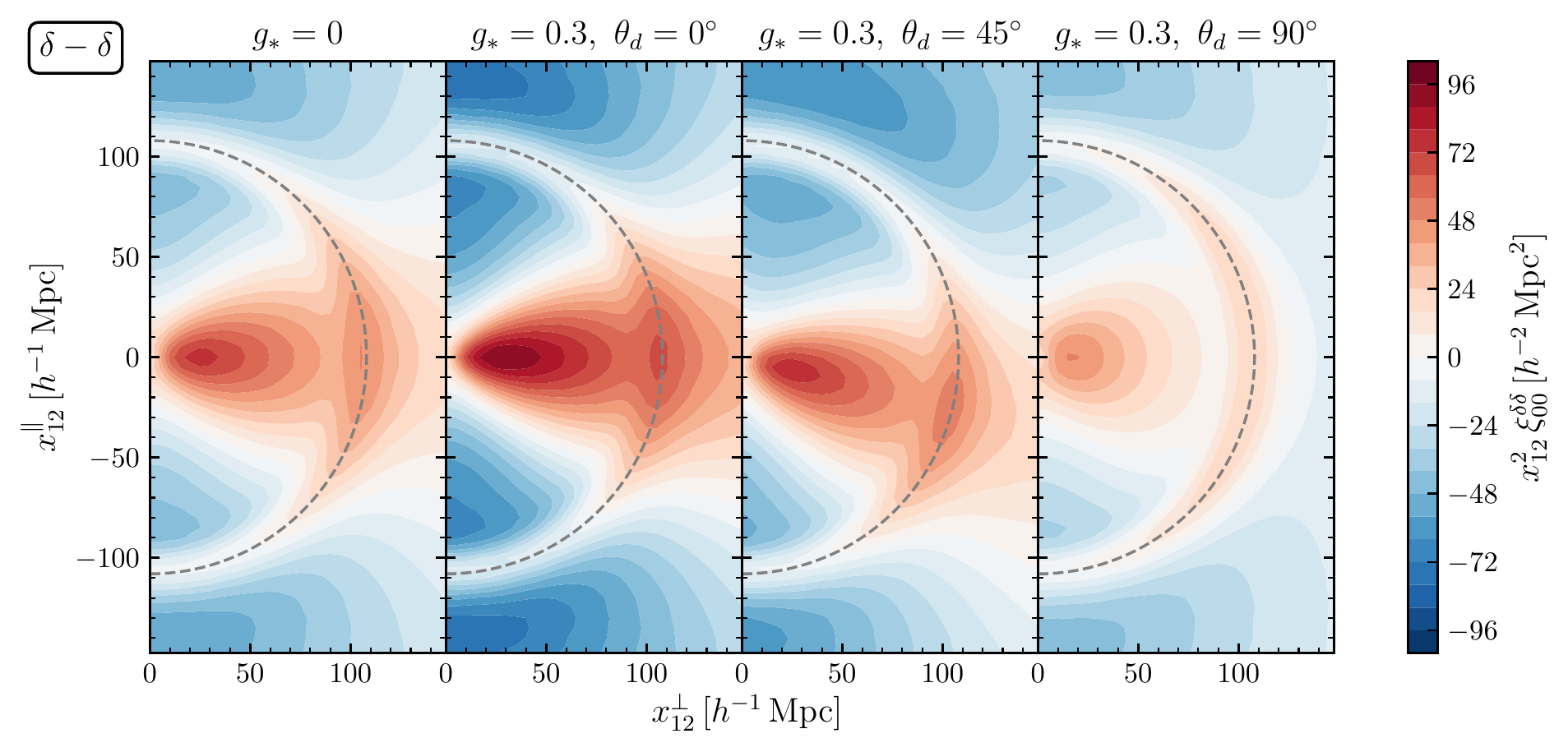}
    \end{center}
  \end{minipage}
  \\
  \begin{minipage}{1.\hsize}
    \begin{center}
      \includegraphics[width=0.93\textwidth]{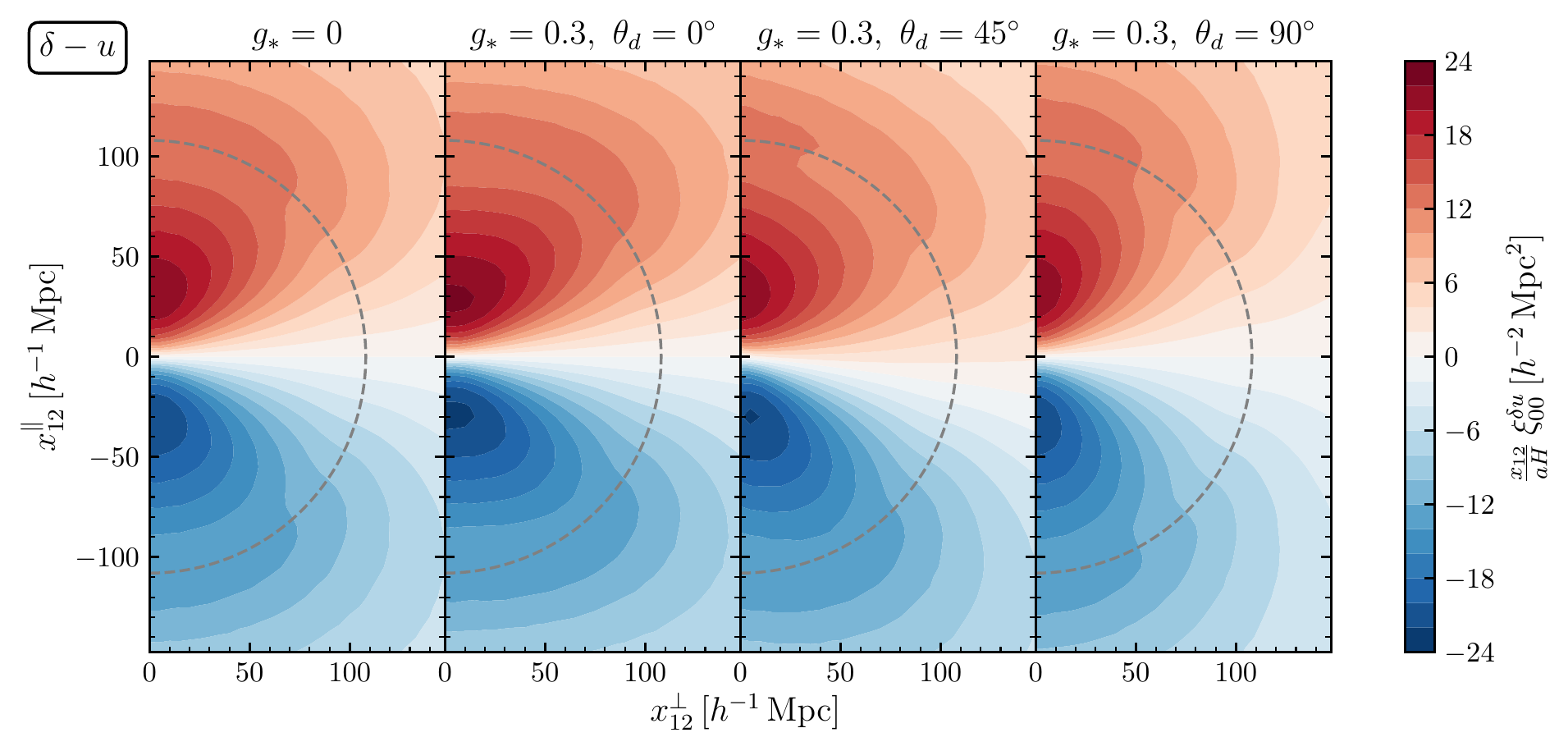}
    \end{center}
  \end{minipage}
  \\
  \begin{minipage}{1.\hsize}
    \begin{center}
      \includegraphics[width=0.93\textwidth]{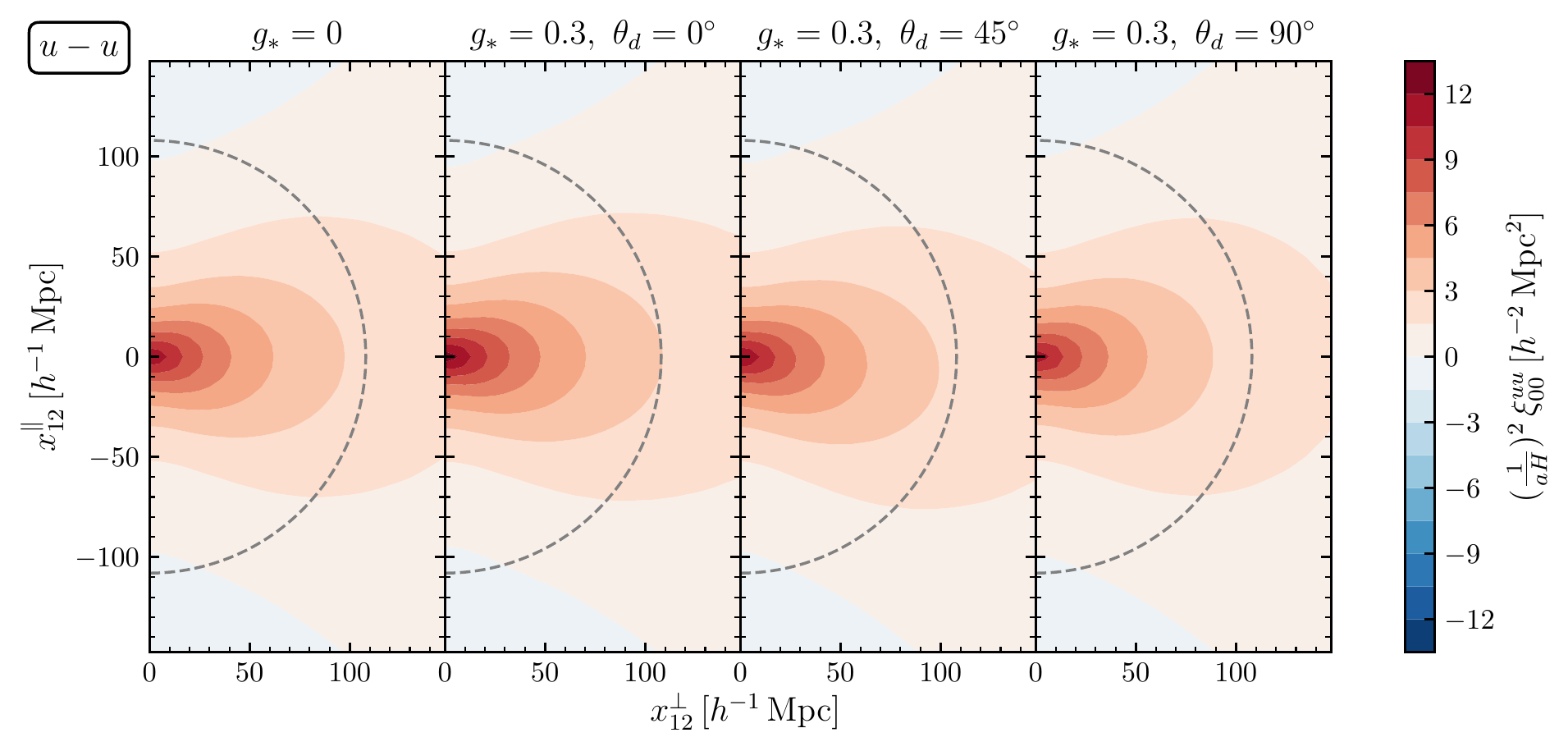}
    \end{center}
  \end{minipage}
  \caption{Same as figure~\ref{fig:xi_flat_2D_dd}, except for the $\delta \delta$ and $\delta u$ correlations in the redshift space and the $uu$ one.}
  \label{fig:xi_flat_2D_dd_dv_vv}
\end{figure}

\begin{figure}[t]
  \begin{center}
    \includegraphics[width=1.\textwidth]{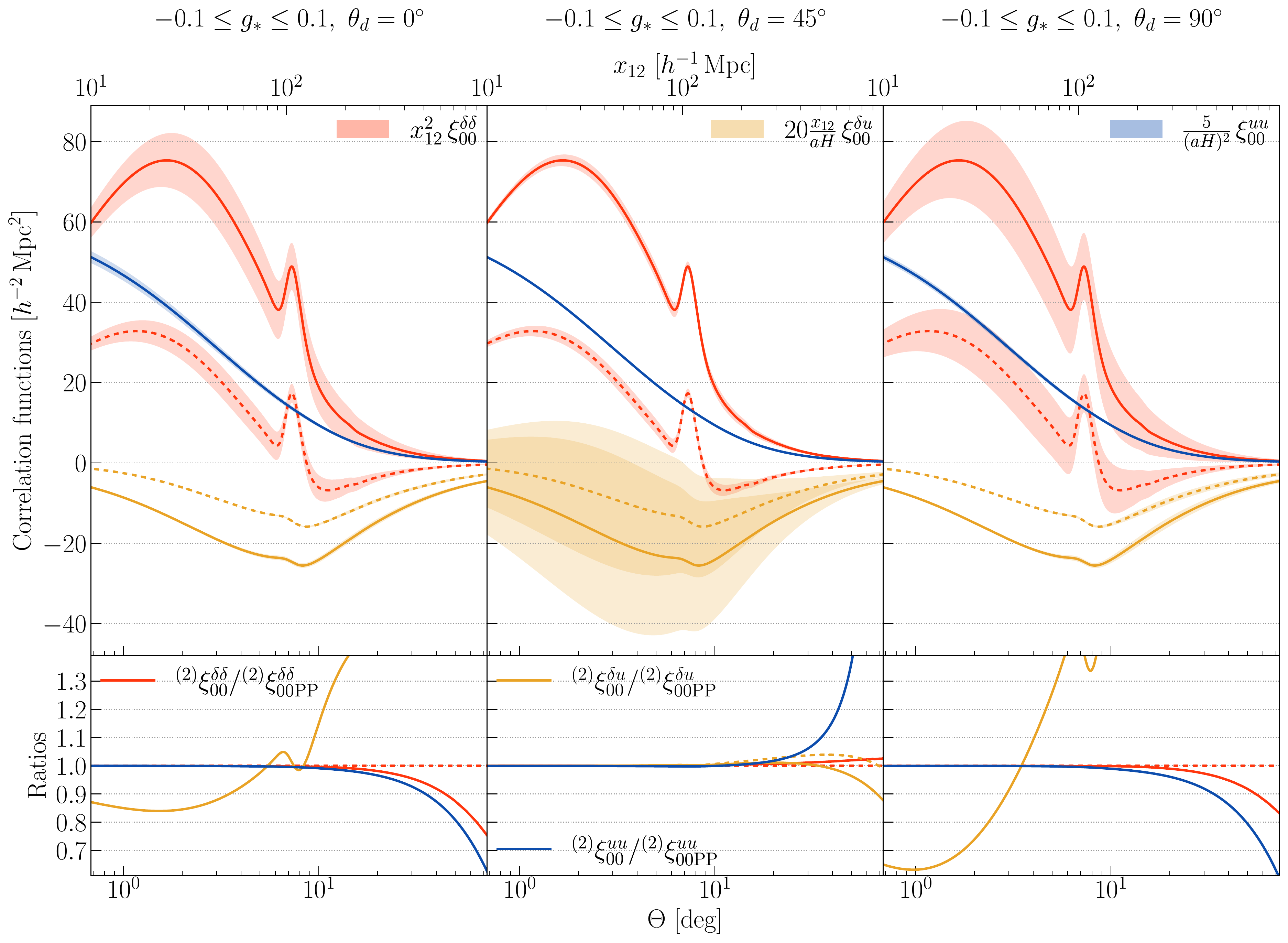}
  \end{center}
  \caption{Same as figure~\ref{fig:xi_per_Theta_dr_vr_rr}, except for the $\delta \delta$, $\delta u$ and $uu$ cases.}
  \label{fig:xi_per_Theta_dd_dv_vv}
\end{figure}

Here we summarize the same results as in section~\ref{sec:result} but for the $\delta \delta$, $\delta u$ and $uu$ cases.

Figure~\ref{fig:xi_flat_2D_dd_dv_vv} describes the intensity distributions of the $\delta \delta$, $\delta u$ and $uu$ correlations in the PP limit on the ($x_{12}^\perp$, $x_{12}^\parallel$) domain for nonzero $g_*$ and several $\theta_d$. One can observe the distortions by nonzero $g_*$ along the $x_{12}^\perp$ axis, the diagonal line and the $x_{12}^\parallel$ axis for $\theta_d = 0^\circ$, $45^\circ$ and $90^\circ$, respectively, similarly to the $\delta \gamma$, $u \gamma$ and $\gamma \gamma$ cases.

Figure~\ref{fig:xi_per_Theta_dd_dv_vv} depicts the $\delta \delta$, $\delta u$ and $uu$ correlations and the ratios between the exact and PP-limit results of their isotropy-breaking parts as a function of $x_{12}$ or $\Theta$ for nonzero $g_*$ and several $\theta_d$. The top panels indicate that the distortion level in the $uu$, $\delta u$ or $\delta \delta$ correlation for a certain $g_*$ varies depending on $\theta_d$, and is maximized for $\theta_d \sim 0^\circ$, $45^\circ$ or $90^\circ$. The bottom panels show that the PP approximation works independently of $\theta_d$ when $\Theta \lesssim 30^\circ$ for the $\delta \delta$ and $u u$ cases, while its validity is not guaranteed even when $\Theta \sim 1^\circ$ for the $\delta u$ case with $\theta_d = 0^\circ$ and $90^\circ$.

\section{Angular correlation functions}\label{appen:Cl}

We here summarize the relation between the configuration-space correlations discussed in the main text and the angular correlations.

The spin-$|\lambda|$ field ${}_{\lambda} X$ can be generally expanded with the spin-weighted spherical harmonics as
\begin{align}
  {}_{\lambda} X({\bf x}) = \sum_{\ell m} {}_{\lambda} a_{\ell m}^{X} \, {}_{\lambda} Y_{\ell m} (\hat{x}) .
\end{align}
Computing the inverse formula: 
\begin{align}
  {}_{\lambda} a_{\ell m}^{X} = \int d^2 \hat{x} \, {}_{\lambda} X({\bf x}) {}_{\lambda} Y_{\ell m}^* (\hat{x}),
\end{align}
with eqs.~\eqref{eq:X}, \eqref{eq:math_Ylm} and \eqref{eq:math_wigner}, we obtain
  \begin{align}
    \begin{split}
  {}_{\lambda} a_{\ell m}^{X} &= {}_{\lambda} \bar{a}_{\ell m}^{X} + {}_{\lambda} a_{\ell m \, \rm ani}^{X \, \rm std} + {}_{\lambda} a_{\ell m \, \rm ani}^{X \, \rm new}, \\
  {}_{\lambda} \bar{a}_{\ell m}^{X} &\equiv 4\pi i^\ell \int \frac{d^3 k}{(2\pi)^3} \bar{\delta}_{\rm m}({\bf k})
  Y_{\ell m}^*(\hat{k})  {}_\lambda {\cal T}_\ell^X(k) ,  \\
  {}_{\lambda} a_{\ell m \, \rm ani}^{X \, \rm std}
  &\equiv 4\pi i^{\ell} 
\int \frac{d^3 k}{(2\pi)^3} \bar{\delta}_{\rm m}({\bf k})
\sum_{\ell ' m'} Y_{\ell' m'}^*(\hat{k})
\frac{1}{2} \sum_{L>0} \sum_M G_{LM}(k)
(-1)^m 
 \begin{pmatrix}
   \ell & \ell' & L \\
   -m & m' & M 
 \end{pmatrix}
 h_{\ell \ell' L}^{000}
 \, {}_\lambda {\cal T}_\ell^X(k) , \\
 {}_{\lambda} a_{\ell m \, \rm ani}^{X \, \rm new}
  &\equiv 4\pi i^\ell \int \frac{d^3 k}{(2\pi)^3} \bar{\delta}_{\rm m}({\bf k})
  \sum_{\ell' m'} Y_{\ell' m'}^*(\hat{k})
   \frac{1}{2}
\sum_{L M}
G_{L M}(k)
(-1)^m
 \begin{pmatrix}
   \ell & \ell' & L \\
   -m & m' & M 
 \end{pmatrix}
     {}_\lambda {\cal S}_{\ell \ell' L}^{X}(k) ,
     \label{eq:alm_X}
    \end{split}
\end{align}
where
\begin{align}
  \begin{split}
  {}_\lambda {\cal T}_{\ell}^{X}(k) &\equiv
  \sum_{j j'}
  \frac{4\pi i^{j - \ell} h_{j  j' \ell}^{0 0 0} h_{j  j'  \ell}^{0 \lambda -\lambda}}{(2\ell + 1)(2j' + 1)}  
  c_{j'}^{X} (k)  j_{j}(kx) , \\
  {}_\lambda {\cal S}_{\ell \ell' L }^{X }(k)
  &\equiv \sum_{j j' J}
 i^{J - \ell} (-1)^{J}
 e_{Ljj'}^{X}
h_{J j' \ell}^{0 \lambda -\lambda} h_{J j \ell'}^{000} 
  \left\{\begin{matrix}
   \ell & \ell' & L \\
   j & j' & J 
  \end{matrix}\right\}
  j_{J}(kx) , 
  \end{split}
\end{align}
and $c_{j}^{X}$ and $e_{Ljj'}^{X}$ are defined in eq.~\eqref{eq:X_coeff}.

Up to linear order of $G_{L>0, M}$, the angular correlation is given as
\begin{align}
\Braket{{}_{\lambda_1} a_{\ell_1 m_1}^{X_1} {}_{\lambda_2} a_{\ell_2 m_2}^{X_2}}
&= \Braket{{}_{\lambda_1} \bar{a}_{\ell_1 m_1}^{X_1} {}_{\lambda_2} \bar{a}_{\ell_2 m_2}^{X_2}}
  + \Braket{{}_{\lambda_1} \bar{a}_{\ell_1 m_1}^{X_1} {}_{\lambda_2} a_{\ell_2 m_2  \, \rm ani}^{X_2 \, \rm std}}
  + \Braket{{}_{\lambda_1} a_{\ell_1 m_1  \, \rm ani}^{X_1 \, \rm std} {}_{\lambda_2} \bar{a}_{\ell_2 m_2}^{X_2}} \nonumber \\
&\quad + \Braket{{}_{\lambda_1} \bar{a}_{\ell_1 m_1}^{X_1} {}_{\lambda_2} a_{\ell_2 m_2  \, \rm ani}^{X_2 \, \rm new}}
  + \Braket{{}_{\lambda_1} a_{\ell_1 m_1  \, \rm ani}^{X_1 \, \rm new} {}_{\lambda_2} \bar{a}_{\ell_2 m_2}^{X_2}} .
\end{align}
Computing this with eqs.~\eqref{eq:alm_X} and \eqref{eq:math_Ylm} leads to
\begin{align}
    \Braket{{}_{\lambda_1} a_{\ell_1 m_1}^{X_1} {}_{\lambda_2} a_{\ell_2 m_2}^{X_2}}
  &= i^{\ell_1 - \ell_2} (-1)^{m_1 + m_2} \sum_{LM }
 \begin{pmatrix}
   \ell_1 & \ell_2 & L \\
   -m_1 & -m_2 & M 
 \end{pmatrix}
     {}_{\lambda_1 \lambda_2} C_{\ell_1 \ell_2}^{L M X_1 X_2} , \label{eq:Cl_X}
\end{align}
where 
\begin{align}
  \begin{split}
{}_{\lambda_1 \lambda_2} C_{\ell_1 \ell_2}^{L M X_1 X_2}
  &=  {}_{\lambda_1 \lambda_2} C_{\ell_1 \ell_2 \, \rm std}^{L M X_1 X_2} + {}_{\lambda_1 \lambda_2} C_{\ell_1 \ell_2 \, \rm new}^{L M X_1 X_2}  ,  \\
{}_{\lambda_1 \lambda_2} C_{\ell_1 \ell_2 \, \rm std}^{L M X_1 X_2}
      &\equiv 
\frac{2}{\pi}  \int_0^\infty k^2 dk 
\bar{P}_{\rm m}(k) G_{LM}(k)
    {}_{\lambda_1} {\cal T}_{\ell_1}^{X_1}(k) {}_{\lambda_2} {\cal T}_{\ell_2}^{X_2}(k)
    h_{\ell_1 \ell_2 L}^{0~0~0} , \\
{}_{\lambda_1 \lambda_2} C_{\ell_1 \ell_2 \, \rm new}^{L M X_1 X_2}
   &\equiv  \frac{2}{\pi}  \int_0^\infty  k^2 dk
   \bar{P}_{\rm m}(k) G_{L M}(k)
   \frac{1}{2}
   \left[  {}_{\lambda_1} {\cal T}_{\ell_1}^{X_1}(k)
     {}_{\lambda_2} {\cal S}_{\ell_2 \ell_1 L}^{X_2}(k)
     + {}_{\lambda_2} {\cal T}_{\ell_2}^{X_2}(k)
     {}_{\lambda_1} {\cal S}_{\ell_1 \ell_2 L}^{X_1}(k)
     \right] .
   \end{split}
\end{align}
Nonzero $G_{L>0, M}$ generated from the isotropy-violating matter power spectrum induce nonzero off-diagonal modes $\ell_1 \neq \ell_2$. The 2PCF $\xi_{\lambda_1 \lambda_2}^{X_1 X_2}$ is related to ${}_{\lambda_1 \lambda_2}C_{\ell_1 \ell_2 }^{LM X_1 X_2}$ as
\begin{align}
  \xi_{\lambda_1 \lambda_2}^{X_1 X_2}({\bf x}_{12}, \hat{x}_1, \hat{x}_2)
  &= \sum_{\ell_1 m_1 \ell_2 m_2}
  i^{\ell_1 - \ell_2} (-1)^{m_1 + m_2} 
  \sum_{LM} 
  \begin{pmatrix}
    \ell_1 & \ell_2 & L \\
    -m_1 & -m_2 & M
  \end{pmatrix} \nonumber \\
  &\quad \times
 {}_{\lambda_1 \lambda_2}C_{\ell_1 \ell_2 }^{LM X_1 X_2}
  {}_{\lambda_1} Y_{\ell_1 m_1} (\hat{x}_1) {}_{\lambda_2} Y_{\ell_2 m_2} (\hat{x}_2) .
\end{align}

The spin-2 ellipticity field ${}_{\pm 2} \gamma$ can be converted into the spin-0 E/B-mode one, whose harmonic coefficient is given by%
\footnote{
See refs.~\cite{Kogai:2018nse,Biagetti:2020lpx} for another convention that is different by sign from ours.
}
\begin{align}
  \begin{split}
  {}_{0} a_{\ell m}^E &\equiv - \frac{1}{2} \left({}_{+2} a_{\ell m}^\gamma + {}_{-2} a_{\ell m}^\gamma \right) , \\
  {}_{0} a_{\ell m}^B &\equiv - \frac{1}{2i} \left({}_{+2} a_{\ell m}^\gamma - {}_{-2} a_{\ell m}^\gamma \right).
  \end{split}
\end{align}
The angular correlation composed of ${}_{0} a_{\ell m}^{E/B}$ also takes the form~\eqref{eq:Cl_X}; and because ${}_{-\lambda} {\cal T}_\ell^X = {}_{\lambda} {\cal T}_\ell^X $ and ${}_{-\lambda} {\cal S}_{\ell \ell' L }^{X } = (-1)^{\ell + \ell'} {}_\lambda {\cal S}_{\ell \ell' L }^{X }$, the following relations hold
\begin{align}
\begin{split}
{}_{00}C_{\ell_1 \ell_2}^{LM, \delta/u, E} 
&= - {}_{0 +2} C_{\ell_1 \ell_2  \, \rm std}^{LM , \delta/u, \gamma}
  - \frac{1 + (-1)^{\ell_1 + \ell_2}}{2} {}_{0 +2} C_{\ell_1 \ell_2 \, \rm new}^{LM, \delta/u, \gamma } , \\
{}_{00}C_{\ell_1 \ell_2}^{LM, \delta/u, B}
   &= - \frac{1 - (-1)^{\ell_1 + \ell_2}}{2i} {}_{0 +2} C_{\ell_1 \ell_2 \, \rm new}^{LM, \delta/u, \gamma } , \\
{}_{00}C_{\ell_1 \ell_2}^{LM EE}
&= {}_{+2 +2} C_{\ell_1 \ell_2 \, \rm std}^{LM \gamma \gamma }
+ \frac{1 + (-1)^{\ell_1 + \ell_2}}{4} 
\left( {}_{+2 +2} C_{\ell_1 \ell_2 \, \rm new}^{LM \gamma \gamma }
+  {}_{+2 -2} C_{\ell_1 \ell_2 \, \rm new}^{LM \gamma \gamma} 
\right) , \\
{}_{00}C_{\ell_1 \ell_2}^{LM EB}
&= 
\frac{1 - (-1)^{\ell_1 + \ell_2}}{4i} 
\left( {}_{+2 +2} C_{\ell_1 \ell_2 \, \rm new}^{LM \gamma \gamma}
-  {}_{+2 -2} C_{\ell_1 \ell_2 \, \rm new}^{LM \gamma \gamma} 
\right) , \\
{}_{00}C_{\ell_1 \ell_2}^{LM BB}
&= 
- \frac{1 + (-1)^{\ell_1 + \ell_2}}{4} 
\left( {}_{+2 +2} C_{\ell_1 \ell_2 \, \rm new}^{LM \gamma \gamma }
-  {}_{+2 -2} C_{\ell_1 \ell_2 \, \rm new}^{LM \gamma \gamma } 
\right) .
\end{split}
\end{align}
These indicate that $\Braket{{}_{0} a_{\ell_1 m_1}^{\delta/u/E} {}_{0} a_{\ell_2 m_2}^{B}}$ and $\Braket{{}_{0} a_{\ell_1 m_1}^{B} {}_{0} a_{\ell_2 m_2}^{B}}$ do not vanish when $\ell_1 + \ell_2 = \rm odd$ and even, respectively.

\section{Useful mathematical identities}\label{appen:math}

The spherical harmonic decompositions: \cite{Shiraishi:2010kd}
\begin{align}
  \begin{split}
    {\cal L}_l(\hat{k} \cdot \hat{x}) &= \frac{4\pi}{2l+1} \sum_m  Y_{lm}(\hat{k}) Y_{lm}^*(\hat{x}), 
    \\
    e^{i{\bf k} \cdot {\bf x}} &= \sum_{lm}
  4\pi i^{l}   j_l(kx) Y_{lm}(\hat{k}) Y_{lm}^*(\hat{x}), \\
    \hat{k} &= \sum_m \boldsymbol{\alpha}^m Y_{1m}(\hat{k}) , \\
    {\bf m}_{\pm}(\hat{x}) &= \pm \sum_m \boldsymbol{\alpha}^m {}_{\mp 1}Y_{1m}(\hat{x}) ,  
        \label{eq:math_expand}
  \end{split} 
\end{align}
where a $m$-dependent vector $\boldsymbol{\alpha}^m$, given by 
\begin{align}
\boldsymbol{\alpha}^m &\equiv \sqrt{\frac{2 \pi}{3}}
  \begin{pmatrix}
   -m (\delta_{m,1}^{\rm K} + \delta_{m,-1}^{\rm K}) \\
   i (\delta_{m,1}^{\rm K} + \delta_{m,-1}^{\rm K}) \\
   \sqrt{2} \delta_{m,0}^{\rm K}
  \end{pmatrix}
,
\end{align}
satisfies $(\boldsymbol{\alpha}^{m})^* = (-1)^m \boldsymbol{\alpha}^{-m}$ and
\begin{align}
  \begin{split}
  \boldsymbol{\alpha}^{m_1} \cdot \boldsymbol{\alpha}^{m_2}
  &= \frac{4 \pi}{3} (-1)^{m_1} \delta_{m_1,-m_2}^{\rm K}, \\
  \sum_m \alpha_i^{m} \alpha_j^{m *} &= \frac{4\pi}{3}  \delta_{ij}^{\rm K}.
  \end{split}
\end{align}

The addition theorem and the orthonormality of the spin-weighted spherical harmonics:
\begin{align}
  \begin{split}
    {}_{s_1} Y_{l_1 m_1}(\hat{n}) {}_{s_2} Y_{l_2 m_2}(\hat{n})
    &= \sum_{s_3 l_3 m_3} {}_{s_3} Y_{l_3 m_3}^*(\hat{n}) 
  h_{l_1 ~ l_2 ~ l_3}^{-s_1 -s_2 -s_3} 
  \begin{pmatrix}
  l_1 & l_2 & l_3 \\
  m_1 & m_2 & m_3 
  \end{pmatrix}
 , 
 \\
 \int d^2 \hat{n} \, {}_{s}Y_{l_1 m_1}(\hat{n}) {}_{s}Y_{l_2 m_2}^*(\hat{n})
 &= \delta_{l_1, l_2}^{\rm K} \delta_{m_1, m_2}^{\rm K} .
 \label{eq:math_Ylm}
 \end{split} 
\end{align}

The addition theorem of the Wigner symbols:
\begin{align}
  \begin{split}
    \frac{\delta_{l_3, l_3'}^{\rm K} \delta_{m_3, m_3'}^{\rm K}}{2l_3+1}
  &=
  \sum_{m_1 m_2}
  \begin{pmatrix}
  l_1 & l_2 & l_3 \\
  m_1 & m_2 & m_3
  \end{pmatrix}
  \begin{pmatrix}
  l_1 & l_2 & l_3' \\
  m_1 & m_2 & m_3' 
  \end{pmatrix} , \\
  \begin{pmatrix}
  l_1 & l_2 & l_3 \\
  m_1 & m_2 & m_3 
  \end{pmatrix} 
\left\{
  \begin{matrix}
  l_1 & l_2 & l_3 \\
  l_4 & l_5 & l_6 
  \end{matrix}
  \right\}
  &=
  \sum_{m_4 m_5 m_6} (-1)^{\sum_{i=4}^6( l_i - m_i) }
  \begin{pmatrix}
  l_5 & l_1 & l_6 \\
  m_5 & -m_1 & -m_6 
  \end{pmatrix} \\
  &\quad \times 
  \begin{pmatrix}
  l_6 & l_2 & l_4 \\
  m_6 & -m_2 & -m_4 
  \end{pmatrix}
  \begin{pmatrix}
  l_4 & l_3 & l_5 \\
  m_4 & -m_3 & -m_5 
  \end{pmatrix}
  , \\
  \begin{pmatrix}
  l_1 & l_2 & l_3 \\
  m_1 & m_2 & m_3 
  \end{pmatrix}
\left\{
  \begin{matrix}
  l_1 & l_2 & l_3 \\
  l_4 & l_5 & l_6 \\
  l_7 & l_8 & l_9
  \end{matrix}
  \right\}
  &=
  \sum_{\substack{m_4 m_5 m_6 \\ m_7 m_8 m_9}} 
  \begin{pmatrix}
  l_4 & l_5 & l_6 \\
  m_4 & m_5 & m_6 
  \end{pmatrix}
  \begin{pmatrix}
  l_7 & l_8 & l_9 \\
  m_7 & m_8 & m_9 
  \end{pmatrix}
  \begin{pmatrix}
  l_4 & l_7 & l_1 \\
  m_4 & m_7 & m_1 
  \end{pmatrix} \\
  & \quad \times 
  \begin{pmatrix}
  l_5 & l_8 & l_2 \\
  m_5 & m_8 & m_2 
  \end{pmatrix}
  \begin{pmatrix}
  l_6 & l_9 & l_3 \\
  m_6 & m_9 & m_3 
  \end{pmatrix} .
  \label{eq:math_wigner}
  \end{split}
\end{align}

The angular integral over a product of unit vectors: \cite{2021arXiv211006732H,Ee:2017jxx}
\begin{align}
  \begin{split}
    \int d^2 \hat{k} \prod_{r=1}^{2n} \hat{k}_{i_r} &= \frac{4\pi}{(2n + 1)!!}
    \left[ \delta_{i_1 i_2}^{\rm K} \delta_{i_3 i_4}^{\rm K} \cdots \delta_{i_{2n-1} i_{2n}}^{\rm K} + \{(2n - 1)!! - 1 \, {\rm perms} \} \right] , \\
    \int d^2 \hat{k} \prod_{r=1}^{2n - 1} \hat{k}_{i_r} &= 0 , \label{eq:math_intvec}
    \end{split} 
  \end{align}
where $n \in \mathbb{N}$.


\bibliography{paper}
\end{document}